\documentclass[12pt]{article}
\usepackage{epsfig}
\usepackage{amsmath,amssymb,amsthm,amscd}
\usepackage{rotating}

\def\IP{{\mathbb P}}

\def\ZZ{{\mathbb Z}}

\def\2{{1\over2}}
\let\<=\langle \let\>=\rangle
\def\new#1\endnew{{\bf #1}}
\def\ifundefined#1{\expandafter\ifx\csname#1\endcsname\relax}
\ifundefined{draftmode}\else    \input draftmode        \fi
\let\Msize=\footnotesize             
\def\BM{\Msize\begin{matrix}}           \def\EM{\end{matrix}}
\def\MN M:#1 #2 N:#3 #4 {{(#1_{#2},#3_{#4})}}
\def\MNH M:#1 #2 N:#3 #4 H:#5,#6 [#7]{{(#1_{#2},#3_{#4})^{#5,#6}_{#7}}}

\newcommand{\ds}{\displaystyle}
\newcommand{\cF}{{\cal F}}

\def\dd{\mathrm{d}}
\def\cI{\mathcal{I}}

\newcommand{\be}{\begin{equation}}
\newcommand{\ee}{\end{equation}}
\newcommand{\bea}{\begin{eqnarray}}
\newcommand{\eea}{\end{eqnarray}}

\def\a{\alpha}

\def\eps{\epsilon}

\begin{document}
\begin{titlepage}
{}~ \hfill\vbox{ \hbox{} }\break

\rightline{hep-th/0612125} \rightline{MAD-TH-06-12} \vskip 1cm

\centerline{\Large \bf Topological string theory on compact
Calabi-Yau:} \vskip 0.2 cm \centerline{\Large \bf modularity and
boundary conditions} \vskip 0.5 cm


\long\def\symbolfootnote[#1]#2{\begingroup%
\def\thefootnote{\fnsymbol{footnote}}\footnote[#1]{#2}\endgroup}

\vskip 30pt \centerline{ {\large \rm Min-xin Huang
\symbolfootnote[1]{minxin@physics.wisc.edu},
Albrecht Klemm\symbolfootnote[7]{aklemm@physics.wisc.edu} and
Seth Quackenbush\symbolfootnote[2]{squackenbush@wisc.edu} }} \vskip .5cm \vskip
30pt
\centerline{$^{\star}$ $^{\star\star}$ $^\dagger${\it Department of Physics, University of
Wisconsin}}
\centerline{$^{\star\star}${\it Department of Mathematics, University of
Wisconsin}}
\centerline{\it Madison, WI 53706, U.S.A.}

\setcounter{footnote}{0}
\vskip 100pt
\begin{abstract}
The topological string partition function $Z(\lambda,t,\bar t)$
$=\exp(\lambda^{2g-2} F_g(t,\bar t))$ is calculated on a compact Calabi-Yau $M$. 
The $F_g(t,\bar t)$ fulfill the holomorphic anomaly
equations, which imply that $\Psi=Z$ transforms as a wave function
on the symplectic space $H^3(M,\mathbb{Z})$. This defines it
everywhere in the moduli space ${\cal M}(M)$ along with preferred
local coordinates. Modular properties of the sections $F_g$ as
well as local constraints from the 4d effective action allow us to
fix $Z$ to a large extent. Currently with a newly found 
gap condition at the conifold, regularity at the orbifold and the most naive bounds
from Castelnuovo's theory, we can provide the boundary data, which
specify $Z$, e.g. up to genus 51 for the quintic.
\end{abstract}

\end{titlepage}
\vfill \eject


\newpage

\baselineskip=16pt

\tableofcontents

\section{Outline}

Coupling topological matter to topological gravity is a key
problem in string theory. Conceptually most relevant is the
topological matter sector of the critical string as it arises e.g. in
Calabi-Yau compactifications. Topological string theory on
non-compact Calabi-Yau manifolds such as ${\cal O}(-3)\rightarrow
\mathbb{P}^2$ is essentially solved either by localization -
\cite{Klemm:1999gm} or large N-techniques~\cite{Aganagic:2003db}
and has intriguing connections to Chern-Simons
theory~\cite{Witten:1992fb}, open-closed string
duality~\cite{Gopakumar:1998ki}, matrix
models~\cite{Dijkgraaf:2002fc}, integrable hierarchies of
non-critical string theory~\cite{Aganagic:2003qj} and 2d
Yang-Mills theory~\cite{Vafa:2004qa}.

However, while local Calabi-Yau manifolds are suitable to study
gauge theories and more exotic field theories in 4d and specific
couplings to gravity, none of the techniques above extends to
compact Calabi-Yau spaces, which are relevant for important
questions in 4d quantum gravity concerning e.g. the properties of
4d black holes~\cite{Ooguri:2004zv} and the wave function in mini
superspace~\cite{Ooguri:2005vr}.

Moreover, while the genus dependence is encoded in the
Chern-Simons and matrix model approaches in a superior fashion by
the ${1\over N^2}$-expansion, the moduli dependence on the
parameter $t$ is reconstructed locally and in a holomorphic limit,
typically by sums over partitions. This yields an algorithm, which
grows exponentially in the world-sheet degree or the
space-time instanton number.

As the total $F_g(t,\bar t)$ are modular invariant sections over
the moduli space ${\cal M}(M)$, they must be generated by a ring
of almost holomorphic modular forms.  This solves the dependence
on the moduli in the most effective way. In the following we will
show that space-time modularity, the holomorphic anomaly equations
of Bershadsky, Cecotti, Ooguri and Vafa, as well as boundary
conditions at various boundary components of the moduli space,
solve the theory very efficiently.

For compact (and non-compact) Calabi-Yau spaces mirror symmetry is
proven at genus zero. The modular properties that we need are also
established  at genus zero. Moreover it has been argued recently
that the  holomorphic anomaly recursions follow from categorical
mirror symmetry \cite{costello,kontsevich}. To establish  mirror
symmetry at higher genus, one needs merely to prove that the same
boundary data fix the $F_g(t,\bar t)$ in the $A$- and the
$B$-model.

\subsection{Extending the Seiberg-Witten approach to gravity}
Seiberg-Witten reconstructed the non-perturbative N=2 gauge gauge
coupling from meromorphic section over ${\cal M}(M)$ using their
modular properties and certain local  data from the effective
action at singular divisors of ${\cal M}(M)$. In~\cite{Huang}~we
reconsidered the problem of topological string on local Calabi-Yau
from the modular point of view and found that the singular
behaviour of the gravitational couplings is restrictive enough to
reconstruct them globally. This can be viewed as the most straightforward 
extension of the  Seiberg-Witten approach to gravitational
couplings.

Note that the problem of instanton counting in these cases is
solved either by geometric engineering, one of the techniques
mentioned above, or more directly by the localisation techniques
in the moduli space of gauge theory instantons by Nekrasov,
Nakajima et. al. It is nevertheless instructive to outline the
general idea in this simple setting\footnote{Maybe the simplest
example of the relation between modularity and the holomorphic
anomaly equations is provided by Hurwitz theory on elliptic curves
\cite{dijkgraafzagier}.}. We focused on the N=2 $SU(2)$
Seiberg-Witten case, but the features hold for any local
Calabi-Yau whose mirror is an elliptic curve with a meromorphic
differential~\cite{Aganagic:2006wq,Huang2}\footnote{With fairly
obvious generalizations for the cases where the mirror is a higher
genus curve. In this case the traditional modular forms of
subgroups $\Gamma$ of $\Gamma_0:=SL(2,\mathbb{Z})$ have to be
replaced by Siegel modular forms of subgroups of
$Sp(2g,\mathbb{Z})$~\cite{Aganagic:2006wq}.} and are as follows
\begin{itemize}
\item The genus $g$ topological string partition functions are given by
\begin{equation}
F^{(g)}(\tau, \bar \tau)=\xi^{2g-2} \sum_{k=0}^{3(g-1)}
\hat E_2^{k}(\tau,\bar \tau) c^{(g)}_k(\tau)\ .
\label{eq:generallocalform}
\end{equation}
Here $\hat E_2(\tau,\bar \tau):= E_2(\tau) + {6 i \over \pi(\bar \tau-\tau)}$ is the
modular invariant anholomorphic extension of the second Eisenstein series
$E_2(\tau)$ and the holomorphic `Yukawa coupling'
$\xi:=C^{(0)}_{ttt}={\partial \tau \over \partial t}$ is an
object of weight $-3$ under the modular $\Gamma\in \Gamma_0=SL(2,\mathbb{Z})$.
For example for pure $N=2$ $SU(2)$ gauge theory $\Gamma=\Gamma(2)$~\cite{Seiberg:1994rs}.
Modular invariance implies then that $c^{(g)}_k(\tau)$ are modular forms of
$\Gamma$ of weight $6(g-1)-2k$.
\item The simple anti-holomorphic dependence of (\ref{eq:generallocalform}) implies
that the only part in $F^{(g)}(\tau, \bar \tau)$ not fixed by the recursive anomaly
equations is the weight $6(g-1)$ holomorphic forms $c^{(k)}_0(\tau)$, which
are finitely generated as a weighted polynomial $c^{(g)}_0(\tau)=p_{6(g-2)}(k_1,\ldots,k_m)$
in the holomorphic generators $G_{k_1},\ldots,G_{k_m}$ of forms of $\Gamma$.
\item The finite data needed to fix the coefficients in $p_{6(g-2)}(k_1,\ldots,k_m)$
are provided in part by the specific leading behaviour of the $F^{(g)}$ at the conifold divisor
\begin{eqnarray} \label{thegap}
F^{(g)}_{\textrm{conifold}}=\frac{(-1)^{g-1}B_{2g}}{2g(2g-2)t_D^{2g-2}}+\mathcal{O}(t_D^0),
\end{eqnarray}
in special local coordinates $t_D$. The order of the leading term
was established in~\cite{BCOV}, the coefficient of the leading
term in \cite{Ghoshal:1995}, and the `gap condition,' i.e. the
vanishing of the following $2g-3$ negative powers in $t_D$
in~\cite{Huang}. This property in particular carries over to the
compact case and we can give indeed a string theoretic explanation
of the finding in~\cite{Huang}.
\item Further conditions are provided by the regularity of the
$F^{(g)}$ at orbifold points in ${\cal M}(M)$. These conditions
unfortunately turn out to be somewhat weaker in the global case
than in the local case.
\end{itemize}

Similar forms as (\ref{eq:generallocalform}) for the $F^{(g)}$ 
appear in the context of Hurwitz theory on elliptic 
curves~\cite{dijkgraafzagier}, of mirror symmetry in $K3$ 
fibre limits\cite{Marino:1998} and on rational 
complex surfaces~\cite{Hosono:1999qc,Hosono:2001gf}.
In the local cases, which have elliptic curves as mirror geometry,
we found~\cite{Huang,Huang2}, that the above conditions
(over)determine the unknowns in $p_{6(g-2)}(k_1,\ldots,k_m)$ and
solve the theory. This holds also for the gauge theories with
matter, which from geometric engineering point of view correspond
to local Calabi-Yau manifolds with several (K\"ahler)
moduli~\cite{Huang2}. Using the precise anholomorphic
dependence and restrictions from space-time modularity one can
iterate the holomorphic anomaly equation with an algorithm which
is exact in the moduli dependence and grows polynomially in
complexity with the genus.

Here we extend this approach further to compact Calabi-Yau spaces
and focus on the class of one K\"ahler moduli Calabi-Yau spaces
$M$ such as the quintic. More precisely we treat the class of one
modulus cases whose mirror $W$ has, parameterized by a suitable
single cover variable, a Picard-Fuchs system  with exactly three
regular singular points: The point of maximal monodromy, a
conifold point, and a point with rational branching. The latter can
be simply a $\mathbb{Z}_d$ orbifold point. This e.g. is the case
for the hypersurfaces where the string theory has an exact
conformal field theory description at this point in terms of an
orbifold of a tensor product of minimal $(2,2)$ SCFT field
theories, the so called the Gepner-model. For some complete
intersections there are massless BPS particles at the branch
locus, which lead in addition to logarithmic singularities.

We find a natural family of coordinates in which the conifold
expansion as well as the rational branched logarithmic
singularities exhibit the gap condition (\ref{thegap}). Despite
the fact that the modular group, in this case a subgroup of ${\rm
Sp}(h^3,\mathbb{Z})$, is poorly understood\footnote{Subgroups of
${\rm Sp}(4,\mathbb{Z})$ in which the monodromy group of the
one-parameter models live have been recently specified
\cite{cyy}.}, we will see that the essential feature carry over to
the compact case. Modular properties, the ``gap condition'',
together with regularity at the orbifold, the leading behaviour
of the $F_g$ at large radius, and Castelenovo's Bound 
determine topological string on one modulus 
Calabi-Yau to a large extent.

\section{The topological B-model}
In this section we give a quick summary of the approach of
\cite{BCOVI,BCOV} to the topological B-model, focusing as fast as
possible on the key problems that need to be overcome: namely the
problem of integrating the anomaly equation efficiently and
the problem of fixing the boundary conditions.

\subsection{The holomorphic anomaly equations}
\label{holomorphicanomaly}
The definition of
$F^{(g)}$ is $F^{(g)}=\int_{ {\cal M}_g} \mu_g$ with measure on ${\cal M}_g$
\begin{equation}
\mu_g=\prod_{i=1}^{3g-3} {\rm d} m_i {\rm d} {\bar m}_{\bar \imath}
\left\langle \prod_{i,\bar \imath}
\int_\Sigma G_{zz} \mu^{(i)\, z}_{\bar z} {\rm d}^2 z
\int_\Sigma G_{\bar z \bar z} \mu^{(i)\, \bar z}_{z} {\rm d}^2 z \right\rangle \ .
\end{equation}
Here the Beltrami differentials $\mu^{(i)\, z}_{\bar z} {\rm d}
{\bar z}$ span $H^1(\Sigma,T\Sigma)$, the tangent space to ${\cal
M}_g$. The construction of the measure $\mu_g$ is strikingly
similar to the one for the bosonic string, once the BRST partner
of the energy-momentum tensor is identified with the
superconformal current $G_{zz}{\rm d } z$ and the ghost number
with the $U(1)$ charge \cite{MB}. $\left\langle \right\rangle$ is
to be evaluated in the internal $(2,2)$ SCFT, but it is easy to
see that it gets only contributions from the topological $(c,c)$
sector.

The holomorphic anomaly equation reads for $g=1$  \cite{BCOVI}
\begin{equation}
{\bar \partial}_{\bar k} \partial_m F^{(1)}={1\over 2}
{\bar C}_{\bar k}^{ij} C^{(0)}_{mij} + \left( {\chi\over 24} -1\right )
G_{\bar km}\ ,
\label{eq:anomalyg1}\
\end{equation}
where $\chi$ is the Euler number of the target space $M$, and for
$g>1$ \cite{BCOV}
\begin{equation}
{\bar \partial}_{\bar k} F^{(g)}={1\over 2}{\bar C}_{\bar k}^{ij}
\left(D_i D_j F^{(g-1)}+ \sum_{r=1}^{g-1}D_i F^{(r)}  D_j F^{(g-r)}\right)\ .
\label{eq:anomaly}
\end{equation}
The right hand side of the equations comes from the complex
co-dimension one boundary of the moduli space of the worldsheet
${\cal M}_g$, which corresponds to pinching of handles. The key idea is  that $\bar \partial_{\bar
k} F^{(g)}=\int_{{\cal M}_g} \bar \partial \partial \lambda_g$,
where $\bar \partial \partial$ are derivatives on ${\cal M}_g$ so
that $\bar \partial_{\bar k} F^{(g)}=\int_{\partial {\cal M}_g}
\lambda_g$. The contribution to the latter integral is  from the
co-dimension one boundary $\partial {\cal M}_g$.

The first equation (\ref{eq:anomalyg1}) can be integrated using
special geometry up to a holomorphic function \cite{BCOVI}, which
is fixed by the consideration in Sect. \ref{boundaryconditions}.

The equations  (\ref{eq:anomaly}) are solved in BCOV using the
fact that due to
\begin{equation}
\bar D_{\bar i} \bar C_{\bar \jmath \bar k \bar l}=\bar D_{\bar \jmath} \bar C_{\bar
\imath \bar k \bar l}
\end{equation}
one can integrate
\begin{equation}
\bar C_{\bar \jmath \bar k \bar l}=e^{-2 K} \bar D_{\bar i}\bar D_{\bar \jmath}
\bar \partial_{\bar k} S
\end{equation}
as
\begin{equation}
S_{\bar \imath} =\bar \partial_{\bar \imath} S,\ \ \
S^j_{\bar \imath} =\bar \partial_{\bar \imath} S^j,\ \ \
\bar C_{\bar k}^{ij}= \partial_{\bar k} S^{ij}\ .
\label{eq:propagators}
\end{equation}
The idea is to write the right hand side of (\ref{eq:anomaly}) as
a derivative w.r.t. $\bar \partial_{\bar k}$. In the first step
one writes
\begin{equation}
\begin{array}{rl}
{\bar \partial}_{\bar k} F^{(g)}=&\ds{{\bar \partial}_{\bar k}
\left({1\over 2}S^{ij}
\left(D_i D_j F^{(g-1)}+ \sum_{r=1}^{g-1}D_i F^{(r)}  D_j F^{(g-r)}\right)\right)} \\[ 3 mm]
&\ds{-{1\over 2} S^{ij}{\bar \partial}_{\bar k}\left(D_i D_j F^{(g-1)}+
\sum_{r=1}^{g-1}D_i F^{(r)}  D_j F^{(g-r)}\right)\ .}
\end{array}
\label{int1:anomaly}
\end{equation}
With the commutator $R^l_{i\bar k j}=-\bar \partial_{\bar k}
\Gamma_{ij}^l=[D_i,\partial_{\bar k}]^l_j= G_{i\bar
k}\delta^l_j+G_{j\bar k}\delta^l_i-C^{(0)}_{ijm} {\bar C}_{\bar
k}^{ml}$ the  second term can be rewritten so that the $\bar
\partial_{\bar k}$ derivative acts in all terms directly on
$F^{(g)}$. Then using (\ref{eq:anomalyg1}, \ref{eq:anomaly}) with
$g'<g$ one can iterate the procedure, which produces an equation
of the form
\begin{equation}
{\bar \partial}_{\bar k} F^{(g)}=\bar \partial_{\bar k}
\Gamma^{(g)}(S^{ij},S^i,S, C^{(<g)}_{i_1,\ldots,i_n})\ ,
\end{equation}
where $\Gamma^{(g)}$ is a functional of  $S^{ij},S^i,S$ and $C^{(<g)}_{i_1,\ldots,i_n}$.
This implies that
\begin{equation}
F^{(g)}=\Gamma^{(g)}(S^{ij},S^i,S, C^{(<g)}_{i_1,\ldots,i_n})+c^{(g)}_0(t)\ ,
\label{fgformal}
\end{equation}
is a solution. Here $c^{(g)}_0(t)$ is the holomorphic ambiguity,
which is not fixed by the recursive procedure. It is holomorphic
in $t$ as well as modular invariant. The {\sl major conceptual
problem} of topological string theory on compact Calabi-Yau is to
find the {\sl boundary conditions} which fix $c^{(g)}_0(t)$. Note
that the problem is not well defined without the constraints from
modular invariance. Using the generalization of the gap condition
in Sect. \ref{boundaryconditions}, the behaviour of the orbifold
singularities in Sect. \ref{quinticorbifold} and Castelnuovo's
bound in Sect. \ref{quinticdbrane} we can achieve this goal to a
large extent.

Properties of the $\Gamma^{(g)}(S^{ij},S^i,S,
C^{(<g)}_{i_1,\ldots,i_n})$ are established using the auxiliary
action
\begin{equation}
Z=\int \dd x \dd \phi \exp(Y+\tilde W)
\label{auxilliaryaction}
\end{equation}
where
\begin{equation}
\begin{array}{rl}
\tilde W(\lambda,x,\phi,t,\bar t)=&\ds{\sum_{g=0}^\infty \sum_{m=0}^\infty
\sum_{n=0}^\infty {1\over m! n!} \tilde C^{(g)}_{i_1,\ldots,i_n,\phi^{m}}
x_{i_1}\ldots x_{i_n} \phi^m}\\
=&\ds{\sum_{g=0}^\infty  \sum_{n=0}^\infty
{\lambda^{2g -2}\over n!} C^{(g)}_{i_1,\ldots,i_n}x_{i_1}\ldots x_{i_n}(1-\phi)^{2-2g-n}}+\left({\chi\over 24}-1\right)
\log\left(1\over {1-\phi}\right) \ ,
\end{array}
\end{equation}
with $C^{(g)}_{i_1,\ldots,i_n}=D_{i_1}\ldots D_{i_n} F^{(g)}$  and
the ``kinetic term'' is given by
\begin{equation}
Y(\lambda , x, \phi;t,\bar t)=-{1\over 2 \lambda^2}
(\Delta_{ij}x^i x^j+ 2 \Delta_{i\phi}x^i \phi +\Delta_{\phi\phi}
\phi^2) + {1\over 2} \log\left(\det \Delta\over \lambda^2\right)\ .
\end{equation}
In \cite{BCOV} it was shown that $\exp(\tilde W)$ fulfills an
equation
\begin{equation}
{\partial \over \partial {\bar t_i}} \exp(\tilde W) =\left[ {\lambda ^2\over 2}
\bar C^{jk}_{\bar \imath} {\partial^2 \over \partial x^j \partial x^k} -
G_{\bar \imath j}  x^j {\partial \over \partial \phi} \right]\exp(\tilde W)\
\label{linearanomaly}
\end{equation}
that is equivalent to the holomorphic anomaly equations, by
checking the coefficients of the $\lambda$ powers, and $\exp(Y)$
fulfills
\begin{equation}
{\partial \over \partial {\bar t_{\bar \imath}}} \exp(Y) =\left[ -{\lambda ^2\over 2}
\bar C^{jk}_{\bar \imath} {\partial^2 \over \partial x^j \partial x^k} -
G_{\bar \imath j}  x^j {\partial \over \partial \phi} \right]\exp(Y) \
                \end{equation}
implying that  $\Delta_{ij}$, $\Delta_{i\phi}$ and $\Delta_{\phi\phi}$ are the inverses
to the propagators $K^{ij}=-S^{ij}$, $K^{i\phi}:=-S^i$ and  $K^{\phi\phi}:=-2S$. A
saddle point expansion of $Z$ gives
\begin{equation}
\log(Z)=\sum_{g=2}^{\infty} \lambda^{2g-2}\left[F^{(g)}-
\Gamma^{(g)}(S^{ij},S^i,S, C^{(<g)}_{i_1,\ldots,i_n})\right]\ ,
\end{equation}
where  $\Gamma^{(g)}(S^{ij},S^i,S, C^{(<g)}_{i_1,\ldots,i_n})$ is
simply the Feynman graph expansion of the action
(\ref{auxilliaryaction}) with the vertices $\tilde
C^{(g)}_{i_1,\ldots,i_n,\phi^{m}}$ and the propagators above.
Moreover it can be easily shown that ${\partial \over
\partial_{\bar t_i}}Z=0$, which implies to all orders that
$F^{(g)}$ can be written as (\ref{fgformal}). This establishes the
reduction of the whole calculation to the determination of the
holomorphic modular invariant sections $c^{(g)}_0(t)\in {\cal
L}^{2g-2}$. However it also reflects the {\sl major technical
problem} in the approach of BCOV, namely that the procedure to
determine the recursive anholomorphic part {\sl grows
exponentially with the genus}. It has been observed in
\cite{Katz:1999} that in concrete cases the terms appearing in the
Feynman graph expansion are not functionally independent. This is a
hint for finitely generated  rings of anholomorphic modular
forms over ${\cal M}(M)$. Using the modular constraints
systematically in each integration step Yamaguchi and Yau
developed a recursive procedure for the quintic whose complexity grows
asymptotically only polynomially, see (\ref{growth}).

Since the B-model is 2d gravity coupled to 2d matter, let us
compare the situation with pure  2d gravity, where the objects of
interest are  correlation functions of $\tau_{d_i}=(2d_i+1)!!
c_1(L_i)^{d_i}$  which are forms on ${\overline {\cal M}_g}$
constructed from the descendent fields
\begin{equation}
F_g(t_0,t_1,\ldots )=\sum_{\{d_i\}} \langle \prod \tau_{d_i}\rangle_g
\prod_{r>0} {t_r^{n_r}\over n_r!}\ .
\end{equation}
Here $\{d_i\}$ are the set of all non-negative integers and  $n_r:={\rm Card}(i:d_i=r)$.

The linear second order differential equations (\ref{linearanomaly}) is the small
phase space analog of  the {\sl Virasoro constraints}
\begin{equation}
L_n Z=0, \qquad n\ge -1
\label{virasoroconstraints}
\end{equation}
on $Z=e^F$ with $F=\sum_{g=0}^\infty \lambda^{2 g-2} F_g$ the free
energy of 2d topological gravity~\cite{wittensurvey}. Indeed the
$L_n$ with $[L_n,L_m]=(n-m) L_{n+m}$ are second order linear
differential operators in the $t_i$. The {\sl non-linear KdV
Hierarchy}, which together with dilaton and string  equation are
equivalent to (\ref{virasoroconstraints})~\cite{wittensurvey}, and
correspond in the small phase space of the B-model to the
holomorphic anomaly equations
(\ref{eq:anomalyg1},\ref{eq:anomaly}). In the $A$-model approaches
to topological string on Calabi-Yau manifolds, such as relative
GW-theory, localisation or attempts to solve the theory via
massive $(2,2)$ models, the descendents are introduced according
to the details of the geometrical construction and then ``summed
away''.

The combinatorial cumbersome information in the descendent sums is
replaced in the B-model by the contraints from the modular group,
holomorphicity and boundary information from the effective 4d
action. As a consequence of this beautiful interplay between
space-time and world-sheet properties one needs only the small
phase space equations (\ref{eq:anomalyg1}, \ref{eq:anomaly},
\ref{linearanomaly}).

This approach requires the ability to relate various local
expansions of $F^{(g)}$ near the boundary of the moduli space.
Sensible local expansions (of terms in the effective action) are
in locally monodromy invariant coordinates. As explained
in~\cite{Aganagic:2006wq} these coordinates in various patches are
related by symplectic transformations on the phase space $H^3(M)$.
The latter extend as metaplectic transformations to the wave
function $\psi=Z$ of the topological string on the
Calabi-Yau~\cite{wittenwavefunction}, which defines the
transformation on the $F^{(g)}$. It will be important for us that
the real
polarisation~\cite{Verlinde:2004ck,Aganagic:2006wq,Gunaydin:2006bz}
defines an unique splitting (\ref{fgformal},\ref{fgdefinition}) of
local expansions of the $F^{(g)}$ in the anholomorphic modular
part determined by the anomaly equations and the holomorphic
modular part $c^{(g)}_0(t)$. Aspects of the wave functions
properties and the various polarizations have been further
discussed
in~\cite{Verlinde:2004ck,Aganagic:2006wq,Gunaydin:2006bz}.

\subsection{Boundary conditions from light BPS states}
\label{boundaryconditions}
Boundaries in the moduli space ${\cal M}(M)$ correspond to degenerations
of the manifold $M$ and general properties of the effective action can be inferred
from the physics of the lightest states. More precisely the light states relevant
to the $F^{(g)}$ terms in the $N=2$ actions are the BPS  states.
Let us first discuss the boundary conditions for $F^{(1)}$ at the singular points
in  the moduli space.
\begin{itemize}
\item At the point of maximal unipotent monodromy in the mirror manifold
$W$, the K\"ahler areas, four, and  six volumes of the original
manifold $M$ are all large. Therefore the lightest string states
are the constant maps $\Sigma_g\rightarrow pt\in M$. For these
Kaluza-Klein reduction, i.e. a zero mode analysis of the A-twisted
non-linear $\sigma$-model is sufficient to calculate the leading
behaviour\footnote{The leading of $F^{(0)}$  at this point is
similarly calculated  and given in (\ref{prepotential}).} of
$F^{(1)}$ as~\cite{BCOVI}
\begin{equation}
F^{(1)}={t_i\over 24} \int c_2 \wedge J_i+ {\cal O}(e^{2 \pi i t})\ .
\label{F1}
\end{equation}
Here   $2 \pi i \,t_i={X^i\over X^0}$ are the canonical K\"ahler parameters,
$c_2$ is the second Chern class, and $J_i$ is the basis for the
K\"ahler cone dual to 2-cycles $C_i$ defining the $t_i:=\int_{C_i} \hat J= \int_{C_i} \sum_{i} t_i J_i$.

\item At the conifold divisor in the moduli space  ${\cal M}(W)$, $W$
develops a nodal singularity, i.e., a collapsing  cycle
with $S^3$ topology. As discussed in sect. \ref{quinticconifold} this
corresponds to the vanishing of the total volume of $M$. The leading
behaviour at this point is universally~\cite{Hosono:1994ax}
\begin{equation}
F^{(1)}= {1\over 12} \log(t_D)+O(t_D)\ .
\label{betaconifold}
\end{equation}
This leading behaviour has been physically explained as the effect of integrating out
a non-perturbative hypermultiplet, namely the  extremal black hole of~\cite{Strominger:1995cz}.
Its mass $\sim t_D$, see (\ref{massbh}), goes to zero at the conifold and it couples to
the $U(1)$ vector in the $N=2$ vectormultiplet, whose lowest component is the modulus $t_D$.
The factor $\frac{1}{12}$ comes from the gravitational one-loop $\beta$-function,  which describes the
running of the $U(1)$ coupling~\cite{Vafa:1995ta}.
A closely related  situation is the one of a shrinking lense space $S/G$.
As explained in \cite{GVbfg} one gets in this case several BPS hyper multiplets as the
bound states of wrapped D-branes, which modifies the  factor $\frac{1}{12}\rightarrow \frac{|G|}{12}$
in the one loop $\beta$-function (\ref{betaconifold}).
\item The gravitational $\beta$-function argument extends also to non-perturbative spectra arising at more
complicated singularities, e.g. with gauge symmetry enhancement and adjoint
matter~\cite{Klemm:1995kj}.
\end{itemize}
For the case of the one parameter families the above boundary information and the fact is
sufficient to fix the holopmorphic ambiguity in $F^{(1)}$.

To learn from  the effective action point of view about the higher
genus boundary behaviour, let us recall that the $F^{(g)}$ as in
$F(\lambda,t)=\sum_{g=1}^\infty \lambda^{2g-2} F^{(g)}(t)$ give
rise to the following term:
\begin{equation}
S^{N=2}_{1-loop}=\int {\rm d}^4 x R_+^2 F(\lambda ,t)\ ,
\label{N=2oneloop}
\end{equation}
where $R_+$ is the self-dual part of the curvature and we identify
$\lambda$ with $F_+$, the self-dual part of the graviphoton field
strength. As explained in \cite{GVI,GVII}, see \cite{MB} for a
review, the term is computed by a one-loop integral in a constant
graviphoton background, which depends only on the left
($SO(4)=SU(2)_L\otimes SU(2)_R$) Lorentz quantum numbers of BPS
particles $P$ in the loop. The calculation is very similar to the
normal Schwinger-loop calculation. The latter computes the
one-loop effective action in an $U(1)$ gauge theory, which comes
from integrating out massive particles $P$ coupling to a constant
background $U(1)$ gauge field. For a self-dual background field
$F_{12}=F_{34}=F$ it leads to the following one-loop determinant
evaluation:
\begin{equation}
S^{S}_{1-loop}=\log\, \det \left(\nabla + m^2 + 2 e \, \sigma_L F\right)=
\int_{\epsilon}^\infty \frac{{\rm d} s}{s} \ds{{ {\rm Tr} (-1)^f \exp({- s m^2 }) \exp({-2 s e \sigma_L F})}\over
4 \sin^2\ds{\left( s e F/2\right)}}\ .
\label{schwingerloop}
\end{equation}
Here the $(-1)^f$ takes care of the sign of the log of the
determinant depending on whether $P$ is a boson or a fermion, and
$\sigma_L$ is the Cartan element in the left Lorentz
representation of $P$. To apply this calculation to the $N=2$
supergravity case one notes, that the graviphoton field couples
to the mass, i.e., we have to identify $e=m$. The loop has two
$R_+$ insertions and an arbitrary, (for the closed string action
even) number, of graviphoton insertions. It turns out \cite{GVII}
that the only supersymmetric BPS states with the Lorentz quantum
numbers
\begin{equation}
\left[\left({\bf\frac{1}{2}},{\bf 0}\right)+2({\bf 0},{\bf 0})\right]\otimes {\cal R}
\label{lorentzrep}
\end{equation}
contribute to the loop. Here  ${\cal R}$ is an arbitrary Lorentz
representation of $SO(4)$. Moreover the two $R_+$ insertions are
absorbed by the first factor in the Lorentz representation
(\ref{lorentzrep}), and the coupling of the particles in the loop
to $F_+$ insertions in the $N=2$ evaluation works exactly  as in
the non-supersymmetric Schwinger-loop calculation above for $P$ in
the representation ${\cal R}$.

What are the microscopic BPS states that run in the loop? They are
related to non-perturbative RR states, which are the only charged
states in the Type II compactification. They come from branes
wrapping cycles in the Calabi-Yau, and as BPS states their masses
are proportional to their central charge (\ref{dbranecharge}). For
example, in the large radius in the type IIA string on $M$, the
mass is determined by integrals of complexified volume forms over
even cycles. E.g., the mass of a 2 brane wrapping a holomorphic
curve ${\cal C}_\beta\in  H^2(M,\mathbb{Z})$ is given by
\begin{equation}
m_\beta=\frac{1}{\lambda} \int_{ {\cal C}_\beta\in H^2(M,\mathbb{Z})} (iJ+B)=
\frac{1}{\lambda} 2 \pi i t\cdot \beta =:\frac{1}{\lambda}t_\beta\ .
\label{massd2}
\end{equation}
We note that $H^2(M,\mathbb{Z})$ plays here the role of the charge
lattice. In the type IIB picture the charge is given by integrals of the
normalized holomorphic $(3,0)$-form $\Omega$. In particular the mass of the extremal
black hole that vanishes at the conifold  is given by
\begin{equation}
m_{BH}=\frac{1}{\lambda \int_{A_D} \Omega} \int_{S^3} \Omega =:\frac{1}{\lambda}t_D\ ,
\label{massbh}
\end{equation}
where $A_D$ is a suitable non-vanishing cycle at the conifold. It
follows from the discussion in previous paragraph that with the
identification $e=m$ and  after a rescaling $s\rightarrow s
\lambda / e$ in (\ref{schwingerloop}), as well as absorbing $F$
into $\lambda$, one gets a result for (\ref{N=2oneloop})
\begin{equation}
F(\lambda, t)= \int_{\epsilon}^\infty \frac{{\rm d} s}{s} \ds{{
{\rm Tr} (-1)^f \exp({- s t }) \exp({-2 s  \sigma_L
\lambda})}\over 4 \sin^2\ds{\left( s \lambda /2\right)}}\ .
\label{n=2loop}
\end{equation}
Here $t$ are the regularized masses, c.f.
(\ref{massd2},\ref{massbh}) of the light particles $P$ that are
integrated out, $f$ is their spins in ${\cal R}$, and $\sigma_L$
is the Cartan element in the representation  ${\cal R}$.

At the large volume point one can related the relevant BPS states
actually to bound states of $D2$  with and infinite tower of $D0$
branes with quantized momenta along the $M$-theory circle.
Moreover the left spin content of the bound state can be related
uniquely to the genus of ${\cal C}_\beta$. This beautiful story
leads, as explained in \cite{GVI,GVII}, after summing over the to
momenta of the $D0$ states  to (\ref{schwingerloopd2d0}), which
together with Castelnuovo's bound for smooth curves leads to very
detailed and valuable boundary information as explained in Sec.
\ref{symplecticinvariants}. It is important to note that all
states are massive so that there are no poles in $F^{(g)}$ for
$g>1$. Hence the leading contribution is regular and can be
extracted from the constant $\beta=0$  contribution in
(\ref{schwingerloopd2d0}) as
\begin{eqnarray} \label{constantmap}
\lim_{t\rightarrow
\infty}F^{(g)}_{\textrm{A-model}}=\frac{(-1)^{g-1}B_{2g}B_{2g-2}}{2g(2g-2)(2g-2)!}\cdot \frac{\chi}{2}\ .
\label{constantmaps}
\end{eqnarray}
The moduli space of constant maps factors into tow components: the
moduli space of the world sheet curve $\Sigma_g$ and the location
of the image point, i.e. $M$. The measures on the components are
$\lambda^{3g-3}$ and $c_3(T_M)$ respectively, explaining the above
formula.

Let us turn to type IIB compactifications near the conifold. As it was checked
with the $\beta$-function  in~\cite{Vafa:1995ta} there is precisely one
BPS hypermultiplet with the Lorentz representation of the
first factor in (\ref{lorentzrep}) becoming massless at the conifold. In this case
the Schwinger-Loop calculation (\ref{n=2loop}) simply becomes
\begin{equation}
F(\lambda, t_D)=\int_{\epsilon}^\infty \frac{{\rm d} s}{s} \ds{{
\exp({- s t_D })} \over 4 \sin^2\ds{\left( s \lambda
/2\right)}}+{\cal O}(t_D^0) =\sum_{g=2}^\infty \left(\lambda\over
t_D\right)^{2g-2} \frac{(-1)^{g-1} B_{2g}}{2 g (2 g-2)}+{\cal
O}(t_D^0)\ . \label{gap}
\end{equation}
Since there are no other light particles, the above equation
(\ref{gap}) encodes all singular terms in the effective action.
There will be regular terms coming from other massive states. This
is precisely the gap condition.

\section{Quintic}
\label{quintic}

We consider the familiar case of the quintic hypersurface in
$\mathbb{P}^4$. The topological string amplitudes $F^{(g)}$ were
computed up to genus 4 in \cite{BCOV, Katz:1999} using the
holomorphic anomaly equation, and fixing the holomorphic ambiguity
by various geometric data. It was also observed a long time ago
\cite{Ghoshal:1995} that the leading terms in $F^{(g)}$ around the
conifold point are the same as $c=1$ strings at the self-dual radius,
thus providing useful information for the holomorphic ambiguity. We
want to explore whether we can find coordinates in which 
the $F^{(g)}$ on the compact Calabi-Yau exhibit the gap structure 
around the conifold point that was recently found for local 
Calabi-Yau geometries~\cite{Huang}. In order to do this, it is useful to 
rewrite the topological string amplitudes
as polynomials \cite{Yamaguchi}. We briefly review the formalism
in \cite{Yamaguchi}.

The quintic manifold $M$ has one K\"ahler modulus $t$ and its mirror
$W$ has one complex modulus $\psi$ and is given by the 
equation~\footnote{Here for later convenience we use a slightly 
different notation from that of
\cite{Yamaguchi}. Their notation is simply related to ours by a
change of variable $\psi^5\rightarrow \psi$.}
\begin{equation}
W=x_1^5+x_2^5+x_3^5+x_4^5+ x_5^5-5\psi^\frac{1}{5}x_1x_2x_3x_4x_5=0\ .
\label{quintic} 
\end{equation} 

There is a relation between $t$ and $\psi$ known as the mirror 
map $t(\psi)$. The mirror map and the genus zero amplitude 
can be obtained using the Picard-Fuchs equation on $W$
\begin{eqnarray}
\{(\psi\partial_{\psi})^4-\psi^{-1}(\psi\partial_{\psi}-\frac{1}{5})
(\psi\partial_{\psi}-\frac{2}{5})(\psi\partial_{\psi}-\frac{3}{5})(\psi\partial_{\psi}-\frac{4}{5})\}\omega=0
\label{differentialoperator}
\end{eqnarray}
We can solve the equation as an asymptotic series around
$\psi\rightarrow \infty$.
\subsection{$\psi=\infty$ expansion and integer symplectic basis}
\label{integralsymplecticbasis}
Here we set the notation for the periods in the integer 
symplectic basis on $W$  and the relation of this basis to the 
D-brane charges on $M$. Eqs. (\ref{periods},\ref{prepotential}) and 
following, apply to all one parameter models.
 
The point $\psi=\infty$  has maximal unipotent monodromy and
corresponds to the large radius expansion of the mirror $M$
\cite{Candelas:1990rm}. In the variable $z={1\over 5^5\psi}$ and using the definitions
\begin{equation}
\omega(z,\rho):= \sum
_{n=0}^{\infty}\frac{\Gamma(5(n+\rho)+1)}{\Gamma^5(n+\rho+1)} z^{n
+\rho}\qquad D^k_\rho \omega:=\left.\frac{1}{(2\pi i)^k k!}\frac{\partial^k}{\partial^k \rho}\omega\right|_{\rho=0}
\end{equation}
one can write  the solutions~\cite{Hosono:1994ax}
\begin{equation}
\begin{array}{rl}
\omega_0&= \omega(z,0)=\sum _{n=0}^{\infty}
\frac{(5n)!}{(n!)^5(5^5\psi)^{n}}\\[ 2mm]
\omega_1&=D_\rho \omega(z,0)={1\over 2 \pi i} \left(\omega_0 \ln(z)+\sigma_1\right)\\[ 2mm]
\omega_2&= \kappa D^2_\rho \omega(z,\rho)-c \omega_0={\kappa\over 2\cdot  (2 \pi i)^2 }\left( \omega_0 \ln^2(z)+2 \sigma_1 \ln(z)+\sigma_2\right) \\[2 mm]
\omega_3&= \kappa D^3_\rho \omega(z,\rho)-c \omega_1+e \omega_0=\frac{\kappa}{6 \cdot  (2 \pi i)^3 }\left( \omega_0 \ln^3(z)+3 \sigma_1 \ln^2(z)+3 \sigma_2\ln(z)+\sigma_3\right)
\end{array}
\label{rawbasis}
\end{equation}
The constants $\kappa,c,e$ are topological intersection numbers,
see below. The $\sigma_k$ are also determined by
(\ref{differentialoperator}). To the first few orders,
$\omega_0=1+120 z+113400z^2+{\cal O}(z^3)$, $\sigma_1=770z+810225
z^2+{\cal O}(z^3)$, $\sigma_2=1150z+{4208175\over 2} z^2+{\cal
O}(z^3)$ and $\sigma_3=-6900z-{9895125\over 2}z^2+{\cal O}(z^3)$.

The solutions (\ref{rawbasis}) can be combined into the period vector
$\vec \Pi$ with respect to an integer symplectic basis\footnote{With $A^i\cap B_j=\delta^i_j$
and zero intersections otherwise.} $(A^i,B_j)$ of $H^3(W,\mathbb{Z})$
as follows \cite{Candelas:1990rm}:
\begin{equation}
\vec \Pi  =
\left(
\begin{array}{c}
\int_{B_1} \Omega\\
\int_{B_2} \Omega\\
\int_{A^1} \Omega\\
\int_{A^2} \Omega
\end{array}
\right)=
\left(
\begin{array}{c}
F_0\\
F_1\\
X_0\\
X_1\\
\end{array}
\right)=
\omega_0\left(
\begin{array}{c}
2{\cal F}^{(0)} -t \partial_t {\cal F}^{(0)}\\
\partial_t {\cal F}^{(0)} \\
1\\
t\\
\end{array}
\right)=\left(
\begin{array}{c}
\omega_3+c\, \omega_1+e\, \omega_0 \\
-\omega_2- a\, \omega_1+c \, \omega_0 \\
\omega_0\\
\omega_1 \\
\end{array}
\right)\ .
\label{periods}
\end{equation}
Physically, $t$ is the complexified area of a degree one curve and is related by the mirror map
\begin{eqnarray}
2 \pi i t(\psi)&=&\int_{\cal C} (iJ+ B)=
{\omega_1\over \omega_0}=-\log(5^5\psi)+\frac{154}{625\psi}+\frac{28713}{390625\psi^{2}}+.. \label{mm}\\
{1\over z}=5^5 \psi&=&\frac{1}{q} + 770 + 421375\,q +
274007500\,q^2 +\ldots \label{imm}
\end{eqnarray}
to $\psi$. In (\ref{imm}), we inverted (\ref{mm}) with  $q=e^{2 \pi i t}$.
The prepotential is given by
\begin{equation}
{\cal F}^{(0)}=-{\kappa\over 3!} t^3-{a\over 2}  t^2+ c t+\frac{e}{2}+ f_{inst}(q) \
\label{prepotential}
\end{equation}
where the instanton expansion $f_{inst}(q)$  vanishes in the large
radius  $q\rightarrow 0$ limit. The constants in
(\ref{periods},\ref{prepotential}) can be related to the classical
geometry of the mirror
manifold~\cite{Hosono:1994ax,Candelas:1990rm}. Denote by  $J\in
H_2(M,\mathbb{Z})$ the K\"ahler form which spans the one-dimensional
K\"ahler cone. Then $\kappa=\int_M J\wedge J\wedge J
$ is the classical triple intersection number, $c={1\over 24}
\int_M c_2(T_M) \wedge J$ is proportional to the  evaluation of
the second Chern class of the tangent bundle $T_M$ against $J$,
$e={\zeta(3)\over (2 \pi i)^3}\int_M c_3(T_M)$ is proportional to
the Euler number, and  $a$ is related to the topology of the divisor
$D$ dual to $J$. $A^1$ is topologically a $T_{\mathbb{R}}^3$,
while analysis at the conifold shows that the dual cycle $B_1$ has
the topology of an $S^3$.

The identification of the central charge formula for compactified
type II supergravity~\cite{Ceresole:1995jg} with the K-theory
charge of $D$-branes as objects $A\in
K(M)$~\cite{Minasian:1997mm,Cheung:1997az},
\begin{equation}
\vec Q\cdot \vec \Pi=-\int_M e^{-\hat J} {\rm ch}(A)\sqrt{{\rm td}( M )}=Z(A),
\label{dbranecharge}
\end{equation}
with $\hat J =t (i J+B)$ for the one-dimensional K\"ahler cone,
checks the $D$-brane interpretation of (\ref{periods})~\cite{Brunner:1999jq,Diaconescu:1999vp,Mayr:2000as}
in the $q\rightarrow 0$ limit on the classical terms $\kappa, a, c$, but misses the  ${\chi \zeta(3)\over (2 \pi i)^3}$
term. Based on their scaling with the area parameter $t$ the periods $(F_0,F_1,X_0,X_1)$
are identified with the masses of $(D_6,D_4,D_0,D_2)$ branes. For smooth intersections and
$D$ the restriction of the hyperplane class we can readily calculate $\kappa,a,c,\chi$
using the adjunction formula, see Appendix A.

\subsection{Polynomial expansion of $F^{(g)}$}
\label{polynomialfg}

>From the periods,  or equivalently the prepotential ${\cal
F}^{(0)}$, we can compute the K\"ahler potential $K:=-\log
(i\left({\bar X}^{\bar a} F_a- X^{a} \bar F_{\bar a}\right))$ and
metric $G_{\psi\bar{\psi}}:=\partial_\psi {\bar\partial}_{\bar
\psi}\, K$ in the moduli space. The genus zero Gromov-Witten
invariants are obtained by expanding ${\cal F}^{(0)}$ in large
K\"ahler parameter in a power series in $q$, see Section
\ref{symplecticinvariants}.

We use the notation of \cite{Yamaguchi} and introduce the following symbols:
\begin{eqnarray}
&&
A_p:=\frac{(\psi\partial_{\psi})^pG_{\psi\bar{\psi}}}{G_{\psi\bar{\psi}}},
~~~B_p:=\frac{(\psi\partial_{\psi})^pe^{-K}}{e^{-K}}, ~~(p=1,2,3,\cdots) \nonumber \\
&& C:=C_{\psi\psi\psi}\psi^3, ~~~ X:=\frac{1}{1-\psi}=:-\frac{1}{\delta}
\label{generators}
\end{eqnarray}
Here $C_{\psi\psi\psi}\sim \frac{\psi^{-2}}{1-\psi}$ is the three
point Yukawa coupling, and $A:=A_1=-\psi\Gamma_{\psi\psi}^{\psi}$
and $B:=B_1=-\psi\partial_\psi K$ are the Christoffel and K\"ahler
connections. In the holomorphic limit $\bar{\psi}\rightarrow
\infty$, the holomorphic part of the K\"ahler potential and the
metric go like
\begin{eqnarray}
e^{-K}\sim \omega_0, ~~~~G_{\psi\bar{\psi}}\sim \partial_\psi t,
\end{eqnarray}
so in the holomorphic limit, the generators $A_p$ and $B_p$ are
\begin{eqnarray}
A_p=\frac{(\psi\partial_{\psi})^p(\partial_\psi t)}{\partial_\psi
t}, ~~~B_p=\frac{(\psi\partial_{\psi})^p \omega_0}{ \omega_0}.
\end{eqnarray}
It is straightforward to compute them in an asymptotic series
using the Picard-Fuchs equation. There are also some derivative
relations among the generators,
\begin{eqnarray}
\psi\partial_\psi A_p=A_{p+1}-AA_p, ~~~ \psi\partial_\psi
B_p=B_{p+1}-BB_p,~~~ \psi\partial_\psi X=X(X-1). \nonumber
\end{eqnarray}
The topological amplitudes in the ``Yukawa coupling frame'' are
denoted as
\begin{eqnarray}
P_g:=C^{g-1}F^{(g)}, ~~~~
P^{(n)}_g=C^{g-1}\psi^{n}C_{\psi^n}^{(g)}.
\label{defPg}
\end{eqnarray}
The A-model higher genus Gromov-Witten invariants are obtained in
the holomorphic limit $\bar{t}\rightarrow \infty$ with a familiar
factor of $\omega_0$ as
\begin{eqnarray} \label{10-14-2.8}
F^{(g)}_{\textrm{A-model}} &=& \lim_{\bar{t}\rightarrow \infty}
\omega_0^{2(g-1)}(\frac{1-\psi}{5\psi})^{g-1}P_g.
\end{eqnarray}
The $P^{(n)}_g$ are defined for $g=0, n\geq 3$, $g=1, n\geq 1$,
and $g=2, n\geq 0$. We have the initial data
\begin{eqnarray}
P^{(3)}_{g=0}&=&1 \nonumber \\
P^{(1)}_{g=1} &=&
-\frac{31}{3}B+\frac{1}{12}(X-1)-\frac{1}{2}A+\frac{5}{3}
\label{Pg01},
\end{eqnarray}
and using the Christoffel and K\"ahler connections we have the
following recursion relation in $n$,
\begin{eqnarray}
P^{(n+1)}_g=\psi\partial_\psi
P^{(n)}_g-[n(A+1)+(2-2g)(B-\frac{1}{2}X)]P^{(n)}_g.
\end{eqnarray}

One can also use the Picard-Fuchs equation and the special
geometry relation to derive the following relations among
generators:
\begin{eqnarray}
B_4&=&2XB_3-\frac{7}{5}XB_2+\frac{2}{5}XB-\frac{24}{625}X \nonumber \\
A_2&=& -4B_2-2AB-2B+2B^2-2A+2XB+XA+\frac{3}{5}X-1.
\end{eqnarray}
By taking derivatives w.r.t. $\psi$ one see that all higher $A_p$
($p\geq 2$) and $B_p$ ($p\geq 4$) can be written as polynomials of
$A$, $B$, $B_2$, $B_3$, $X$. It is convenient to change variables
from $(A,B,B_2,B_3,X)$ to $(u,v_1,v_2,v_3,X)$ as
\begin{eqnarray}
&& B=u, ~~~ A=v_1-1-2u,~~~ B_2=v_2+uv_1, \nonumber \\
&& B_3=v_3-uv_2+uv_1X-\frac{2}{5}uX.
\end{eqnarray}
Then the  main result of \cite{Yamaguchi} is the following
proposition.
\\

\noindent{\bf Proposition}: Each $P_g$ ($g\geq 2$) is a degree
$3g-3$ inhomogeneous polynomial of $v_1$, $v_2$, $v_3$, $X$, where
one assigns the degree $1,2,3,1$ for $v_1, v_2, v_3, X$,
respectively.
\\

For example, at genus two the topological string amplitude is
\begin{eqnarray}
P_2 &=& \frac{25}{144} - \frac{625}{288} v_1 + \frac{25}{24} v_1^2
- \frac{5}{24}v_1^3 - \frac{625}{36}v_2 + \frac{25}{6}v_1v_2 +
\frac{350}{9}v_3 - \frac{5759}{3600}X  \nonumber \\ && -
\frac{167}{720}v_1X + \frac{1}{6}v_1^2X - \frac{475}{12}v_2X +
\frac{41}{3600}X^2- \frac{13}{288}v_1X^2 + \frac{X^3}{240}.
\end{eqnarray}
The expression for the $P_g$, $g=1,\ldots,12 $ for all models
discussed in this paper can be obtained in a
Mathematica readable form on \cite{webpage}, see ``Pgfile.txt.''

\subsection{Integration of the holomorphic anomaly equation}
\label{integration}
The anti-holomorphic derivative $\partial_{\bar{\psi}}B_p$ of
$p\geq 2$ can be related to $\partial_{\bar{\psi}}B$
\cite{Yamaguchi}. Assuming $\partial_{\bar{\psi}}A$ and
$\partial_{\bar{\psi}}B$ are independent, one obtains two
relations for $P^{(n)}_g$ from the BCOV holomorphic anomaly
equation as the following:
\begin{eqnarray}
\frac{\partial P_g}{\partial u}&=&0 \label{BCOV1} \\
(\frac{\partial}{\partial{v_1}}-u\frac{\partial }{\partial
{v_2}}-u(u+X)\frac{\partial}{\partial{v_3}})P_g&=&-\frac{1}{2}(P^{(2)}_{g-1}+\sum_{r=1}^{g-1}P^{(1)}_rP^{(1)}_{g-r})
\label{BCOV2}
\end{eqnarray}
The first equation (\ref{BCOV1}) implies there is no $u$
dependence in $P_g$, as already taken into account in the main
proposition of \cite{Yamaguchi}. The second equation (\ref{BCOV2})
provides a very efficient way to solve $P_g$ recursively up to a
holomorphic ambiguity. To do this, one writes down an ansatz for
$P_g$ as a degree $3g-3$ inhomogeneous polynomial of $v_1$, $v_2$,
$v_3$, $X$, plugs in equation (\ref{BCOV2}), and uses the lower
genus results $P_r^{(1)}$, $P^{(2)}_r$ ($r\leq g-1$) to fix the
coefficients of the polynomials. The number of terms in general
inhomogeneous weighted polynomials in $(v_1$, $v_2$, $v_3,X)$ with
weights $(1,2,3,1)$ is given by the generating function
\begin{equation}
{1\over {(1-x)^3 (1-x^2) (1-x^3)}}=\sum_{n=0}^\infty p(n) x^n\ .
\end{equation}
It is easy to see from the equation (\ref{BCOV2}) that the terms
$v_1,\ldots v_1^{2g-4}$ vanish, which implies that the number of
terms in  $P_g$ is
\begin{equation}
n_g= p(3 g-3)-(2 g-4),
\end{equation}
e.g. $n_g= 14, 62, 185, 435, 877, 1590, 2666, 4211, 6344$ for
$g=1,\ldots,10$. Comparing with $\sum_{n=1}^\infty \tilde p(n)
x^n={1\over (1-x)^5}$ it follows in particular that
asymptotically
\begin{equation}
n_g \preceq (3g-3)^4 \ . \label{growth}
\end{equation}

Note that equation (\ref{BCOV2}) determines the term
$\hat  P_g(v_1,v_2,v_3,X)$ in
\begin{equation}
P_g=:\hat P_g(v_1,v_2,v_3,X) + f^{(g)}(X)\
\label{fgdefinition}
\end{equation}
completely. We can easily understand why the terms in the modular
as well as holomorphic ambiguity $f^{(g)}(X)$ are not fixed, by
noting that the equation (\ref{BCOV2}) does not change when we add
a term proportional to $X^i$ to $P_g$. This ambiguity must have
the form~\cite{Katz:1999,Yamaguchi}
\begin{eqnarray} \label{10-14-2.16}
f^{(g)}=\sum_{i=0}^{3g-3}a_iX^i\ .
\end{eqnarray}
The maximal power of $X$ is determined by (\ref{gap}). This
follows from (\ref{defPg}) and the universal behaviour of $t_D\sim
\delta+ {\cal O}(\delta^2)$ at the conifold (\ref{dualmirrormap}).
Note in particular from (\ref{leadingconifold}) that $\hat  P_g$
is less singular at that point. The minimal power in
(\ref{10-14-2.16}) follows from (\ref{constantmaps}) for $\psi
\sim \infty$ and the leading behaviour of the solutions in Sect.
\ref{integralsymplecticbasis}.

We will try to fix these $3g-2$ unknown constants $a_i$
($i=0,1,2\cdots 3g-3$) by special structure of expansions of
$F^{(g)}$ around the orbifold point $\psi\sim 0$ and the conifold
point $\psi\sim 1$. Before proceeding to this, we note the
constant term is fixed by the known leading coefficients in large
complex structure modulus limit $\psi\sim \infty$ in
\cite{Marino:1998, GVII, FP}.

The leading constant terms in the A-model expansion $\psi\sim
\infty$ come from the constant map from the worldsheet to the
Calabi-Yau (\ref{constantmap}). The large complex structure
modulus behavior of $X$ is
\begin{eqnarray}
X\sim \frac{1}{\psi}\sim q=e^{2\pi it}.
\end{eqnarray}
So only the constant term $a_0$ in the holomorphic ambiguity
contributes to leading term in A-model expansions
(\ref{constantmap}), and is thus fixed. We still have $3g-3$
coefficients $a_i$ ($i=1,2,\cdots, 3g-3$) to be fixed.

\subsection{Expansions around the orbifold point $\psi=0$}
\label{quinticorbifold}
To analyze the $F^{(g)}$ in a new region of the 
moduli space we have to find the right choice of 
polarization. To do this we analytically 
continue the periods to determine the symplectic 
pairing in the new region and pick the new 
choice of conjugated varaibles.      

The solutions of the  Picard-Fuchs equation around the orbifold point$\psi\sim 0$
are four power series solutions with the indices
$\frac{1}{5},\frac{2}{5},\frac{3}{5},\frac{4}{5}$
\begin{equation}
\begin{array}{rl}
\omega_{k}^{\rm orb}&=\ds{\psi^{\frac{k}{5}}\sum_{n=0}^\infty
\frac{\left(\left[\frac{k}{5}\right]_n\right)^5}{\left[k\right]_{5 n}}\left(5^5 \psi\right)^n} \\[2 mm]
&=\ds{-\frac{\Gamma(k)}{\Gamma^5\left(\frac{k}{5}\right)}\int_{{\cal
C}_0} \frac{\dd s}{e^{2 \pi i s}-1}
\frac{\Gamma^5\left(s+\frac{k}{5}\right)}{\Gamma(5 s+k)}
\left(5^5 \psi\right)^{s+\frac{k}{5}} \ , \qquad k=1,\ldots, 4}\ .
\label{orbifoldbasis}
\end{array}
\end{equation}
\parbox{14cm} 
{

   \begin{center}
   \mbox{
             \epsfig{file=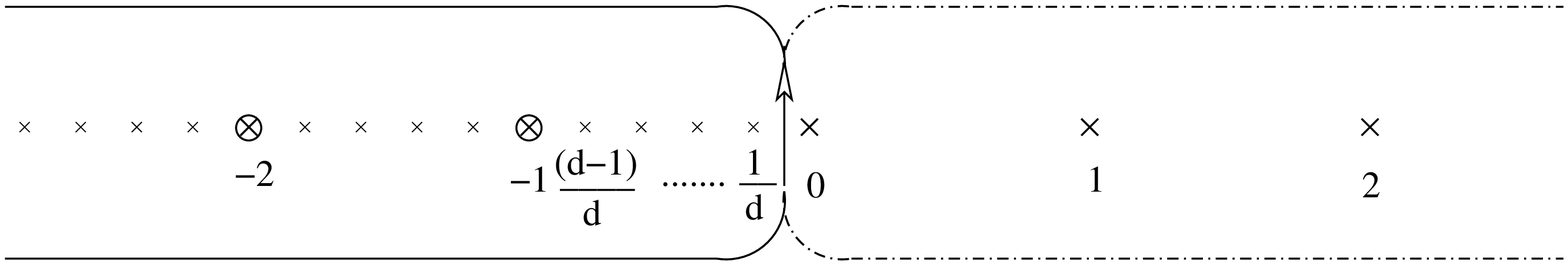,width=14cm}
}
   \end{center}
}
\vskip -2mm
\centerline{ $\qquad\qquad$ ${\cal C}_\infty$ for $|\psi|> 1$ $\qquad\qquad\qquad$ ${\cal C}_0$ for $|\psi|< 1$.}
\vskip 2mm

The Pochhammer symbol are defined as $[a]_n:=\frac{\Gamma(a+n)}{\Gamma(a)}$ and we
normalized the first coefficient in  $\omega_{k}^{\rm orb}= \psi^\frac{k}{5}+{\cal O}(\psi^\frac{5+k}{5})$
to one.  The expression in the first line is recovered from the integral representation
by noting that the only poles inside ${\cal C}_0$ for which the integral
converges for $|\psi|<0$ are from $g(s)=\frac{1}{\exp(2 \pi i s)-1}$, which behaves
at $s^\eps_{-n}=n-\epsilon$, $n\in \mathbb{N}$ as $g(s^\eps_{-n})\sim-\frac{1}{2\pi i \epsilon}$.

Up to normalization this basis of solutions is canonically
distinguished, as it diagonalizes the $\mathbb{Z}_5$ monodromy at
$\psi=0$. Similar as for the $\mathbb{C}^3/\mathbb{Z}_3$
orbifold~\cite{Aganagic:2006wq}, it can be viewed as a twist field
basis. Here this basis is induced from the twist field basis of
$\mathbb{C}^5/\mathbb{Z}_5$. As it was argued
in~\cite{Aganagic:2006wq} for $\mathbb{C}^3/\mathbb{Z}_3$, this
twist field basis provides the natural coordinates in which the
$F^{(g)}$ near the orbifold point can be interpreted as generating
functions for orbifold Gromov-Witten invariants. Following up on
foundational work on orbifold Gromov-Witten theory \cite{CR} and
examples in two complex dimensions~\cite{BGP} this prediction has
been checked by direct computation of orbifold  Gromov-Witten
invariants~\cite{CCIT} at genus zero. This provides a beautiful
check on the global picture of mirror symmetry.

As explained in~\cite{Aganagic:2006wq} the relation between the
large radius generating function of Gromov-Witten invariants and
generating function of orbifold Gromov-Witten invariants is
provided by the metaplectic transformation of the wave
function~\cite{wittenwavefunction}. Since the change of the phase
space variables from large radius to the orbifold the symplectic
form is only invariant up to scaling one has to change the
definition of the string coupling, which plays the role of
$\hbar$ in the  metaplectic transformation.  These can be viewed
as small phase space specialization of the  metaplectic
transformation on the large phase space~\cite{GiventhalCoates}.

Using the modular invariance of the anholomorphic $\hat F^g(t,\bar
t)$ it has been further shown in~\cite{Aganagic:2006wq} that this
procedure of obtaining the transformed holomorphic wave function
is simply equivalent to taking the holomorphic limit on the $\hat
F^g(t,\bar t)$. This point was made~\cite{Aganagic:2006wq} for the
local case. But the only point one  has to keep in mind for the
global case is that $\hat F^g(t,\bar t)$ are globally well defined
sections of the K\"ahler line bundle. i.e. one has to perform a
K\"ahler transformation along with the holomorphic limit.

We will now study  the transformation from the basis
(\ref{periods}) to the basis (\ref{orbifoldbasis}) to make B-model
prediction along the lines of~\cite{Aganagic:2006wq} with the
additional K\"ahler transformation. Since the symplectic form
$\omega$ on the moduli space is  invariant under monodromy and
$\omega_k^{orb}$ diagonalizes the $\mathbb{Z}_5$ monodromy, we
must have in accordance with the expectation from the orbifold
cohomology $H^*(\mathbb{C}^5/\mathbb{Z}_5)=\mathbb{C}{\bf
1}_0\oplus\mathbb{C}{\bf 1}_1\oplus\mathbb{C}{\bf
1}_2\oplus\mathbb{C} {\bf 1}_3\oplus\mathbb{C}{\bf 1}_4$
\begin{equation}
\omega=\dd F_k\wedge \dd X_k= -\frac{5^4s_1}{6}\dd
\omega^{orb}_4\wedge \dd \omega^{orb}_1+ \frac{5^4s_2}{2}\dd
\omega^{orb}_3\wedge \dd \omega^{orb}_2\ .
\end{equation}
The rational factors above have been chosen to match constraints
from special geometric discussed below.  Similarly the monodromy
invariant K\"ahler potential must have the form
\begin{equation}
e^{-K}=\sum_{k=1}^4 r_k \omega_k^{\rm orb} \overline{\omega_k^{\rm orb}}\ .
\label{eK}
\end{equation}
To obtain the $s_i,r_i$ by analytic continuation to the basis (\ref{periods}) we follow \cite{Candelas:1990rm} for the
quintic and the generalisation in~\cite{Klemm:1992tx} for other cases and  note that the integral converges for
$|\psi|>1$ due to the asymptotics of the $f(s)=\frac{\Gamma^5(s+k/5)}{\Gamma(5s+k)}$ term,
when the integral is closed along ${\cal C}_\infty$~\cite{Candelas:1990rm}. At $s^\eps_{n}=-n-\epsilon$
the  $g(s^\eps_{n})$ pole is compensated by the  $f(s^\eps_{n})$ zero and at
$s^\eps_{n,k}= -n-k/5-\epsilon$ we note the expansions
\begin{equation}
\begin{array}{rl}
g(s^\eps_{n,k})=&{\frac{\a^k}{1-\a^k}+
\frac{2\pi i\a^k}{(1-\a^k)^2}\eps+
\frac{(2\pi i)^2\a^k(1+\a^k)}{2(1-\a^k)^3}\eps^2+
\frac{(2\pi i)^3\a^k(1+4\a^k+\a^{2k})}{6(1-\a^k)^4}\eps^3+{\cal O}(\eps^4)}\\[3 mm]
f(s^\eps_{n,k})=&{\frac{\kappa \omega_0(n)}{\eps^4}+
\frac{\kappa \sigma_1(n)}{\eps^3}+
\frac{1}{\eps^2}\left(\frac{\kappa \sigma_2(n)}{2}+\frac{(2\pi i)^2c_{2J} \omega_0(n)}{24}\right)+}\\[3 mm]
&{\frac{1}{\eps}\left(\frac{\kappa \sigma_3(n)}{6}+\frac{(2\pi i)^2 c_{2J}\sigma_1(n)}{24}+\chi \zeta(3)
\omega_0(n)\right)+{\cal O}(\eps^0)}\\[3 mm]
\left(5^5\psi\right)^{s^\eps_n}=&{z^n(1+\log(z)\eps+\frac{1}{2}\log(z)^2\eps^2+\frac{1}{6}\log(z)^3\eps^3+{\cal O}(\eps^4)})
\end{array}.
\end{equation}
Here  $\alpha=\exp(2 \pi i/5)$. $\kappa=\int_MJ^3$,
$c_{2J}=\int_M c_2 J$, and $\chi=\int_M c_3$ are calculated
in App. \ref{intersection}~\footnote{In fact using the
generalization of (\ref{orbifoldbasis}) in~\cite{Klemm:1992tx} it
is easily shown that the combinatorics, which leads to
(\ref{aconst}), are the same as the ones leading to the occurrence
of the classical intersections here.}.  The $\omega_0(n)$,
$\sigma_i(n)$ are coefficients of the series we encountered in
sec. \ref{integralsymplecticbasis}. Performing the residue
integration and comparing with (\ref{rawbasis},\ref{periods}) we
get
\begin{equation}
\omega_k^{orb}=\frac{(2 \pi i)^4 \Gamma(k)}{\Gamma^5\left(\frac{k}{5}\right)}\left(
\frac{\a^k F_0}{1-\a^k}-
\frac{\a^k F_1}{(1-\a^k)^2}+
\frac{5\a^k(\a^{2 k}-\a^k +1)X_0}{(1-\a^k)^4}+
\frac{\a^k(8\a^{k}-3)X_1}{(1-\a^k)^3}\right)
\label{matoi}
\end{equation}
It follows with $r_i=\frac{\Gamma^{10}\left(\frac{k}{5}\right)}{\Gamma^2(k)} c_i$
that
\begin{eqnarray}
c_1&=&-c_4=\a^2 (1-\a)(2+\a^2+\a^3), \quad  c_3=-c_2=\a(2+\a-\a^2-2 \a^3)\nonumber \\
s_1&=&s_2 =-\frac{1}{5^5(2\pi i)^3} \label{metaplectic}\\
\left(\begin{array}{c}F_0\\ F_1\\X_0\\X_1 \end{array} \right)&=&
\psi^{1/5}\frac{\a \Gamma^5\left(1\over 5\right)}{(2 \pi i)^4} \left(\begin{array}{c}
(1-\a)(\a-1-\a^2)\\
\frac{1}{5}(8-3 \a) (1-\a)^2\\
(1-\a+\a^2)\\
\frac{1}{5}(1-\a)^3 \end{array}\right)+{\cal O}(\psi^{2/5})\ . \label{leading}
\end{eqnarray}
Eq. (\ref{metaplectic}) implies that up to a rational rescaling of
the orbifold periods the transformation of the wave functions from
infinity to the orbifold is given by a metaplectic transformation
with the same rescaling of the string coupling as for the
$\mathbb{C}^3/\mathbb{Z}_3$ case in \cite{Aganagic:2003db}. Eq.
(\ref{leading}) implies that there are no projective coordinates
related to an  ${\rm Sp}(4,\mathbb{Z})$ basis, which would vanish
at the orbifold. This means that there is no massless RR state in
the K-theory charge lattice which  vanishes at the orbifold point.
We note further that after rescaling of the orbifold periods the
transformation (\ref{matoi}) can be chosen to lie in ${\rm
Sp}(4,\mathbb{Z}[\a,\frac{1}{5}])$.

We can define the analogue of mirror map at the orbifold point,
\begin{eqnarray}
s=\frac{\omega_2^{orb}}{\omega^{orb}_1}=\psi^{\frac{1}{5}}(1+\frac{13\psi}{360}+\frac{110069\psi^{2}}{9979200}
+\mathcal{O}(\psi^{3}))
\end{eqnarray}
where  we use the notation $s$, as in~\cite{Aganagic:2006wq}, to
avoid confusion with the mirror map in the large volume limit. We
next calculate the genus zero prepotential at the orbifold point.
For convenience let us rescale our periods $\hat \omega_{k-1}=
5^{3/2} \omega^{orb}_{k}$. The Yukawa-Coupling is transformed to
the $s$ variables as
\begin{equation}
C_{sss}=\frac{1}{\hat \omega_0^2}
\frac{5}{\psi^2(1-\psi)} \left(\frac{\partial \psi}{\partial s}\right)^3=
5 +\frac{5}{3} s^5 +\frac{5975}{6048} s^{10}+\frac{34521785}{54486432} s^{15}+{\cal O}(s^{20})  \ .
\end{equation}
A trivial consistency check of special geometry is that the genus
zero prepotential $F^{(0)}=\int \dd s \int \dd s \int \dd s \
C_{sss}$ appears in the periods $\hat \Pi^{orb}=( \hat
\omega_0,\hat \omega_1,\frac{5}{2!} \hat \omega_2,-\frac{5}{3!}
\hat \omega_3)^T$ as
\begin{equation}
\hat \Pi^{orb}=\hat \omega_0 \left(\begin{array}{c}
1\cr
s\cr
\partial_s F_{\textrm{A-orbf.}}^{(0)}\cr
2 F_{\textrm{A-orbf.}}^{(0)}- s\partial_s F_{\textrm{A-orbf.}}^{(0)}
\end{array}
\right)\ .
\end{equation}
This can be viewed also as a simple check on the lowest order 
meta-plectic transformation of $\Psi$ which is just the Legendre 
transformation. Note that the Yukawa coupling is 
invariant under the $\mathbb{Z}_5$ which acts as
$s\mapsto \alpha s$. $\mathbb{Z}_5$ implies further that there can be no
integration constants, when passing from $C_{sss}$ to $F_0$ and the coupling
$\lambda$ must transform with $\lambda \mapsto \alpha^\frac{3}{2} \lambda$ to
render $F(\lambda,s,\bar s)$ invariant.

The holomorphic limit  $\bar{\psi}\rightarrow 0$  of K\"ahler
potential and metric follows from (\ref{eK}) by extracting the
leading anti-holomorphic behaviour. Denoting\footnote{This  is to
make contact with the other one modulus cases. Of course if
$a_1=a_2$ a log singularity appears and the formula does not
apply.} by $a_k$ the leading powers of  $\omega_k^{orb}$  we find
\begin{equation}
\lim_{\bar \psi\rightarrow 0} e^{-K}=r_1 \bar \psi^{a_1}
\omega^{orb}_1, \qquad \lim_{\bar \psi\rightarrow 0} G_{\psi\bar
\psi}= \bar \psi^{a_2-a_1-1}
\frac{r_2}{r_1}\left(\frac{a_2}{a_1}-1\right) \frac{\partial
s}{\partial \psi} \ .
\end{equation}
Note that the constants and  $\bar \psi$ and its leading power
are irrelvant for the holomorphic limit of the generators
(\ref{generators})
\begin{eqnarray}
X&=& \frac{1}{1-\psi}=1+\psi+\psi^{2}+\mathcal{O}(\psi^{3})
\nonumber \\ A&=& -\frac{4}{5}+\frac{13}{60}\psi+
\frac{3551}{18144}\psi^2+\mathcal{O}(\psi^{3}) \nonumber \\
B&=&
\frac{1}{5}+\frac{1}{120}\psi+\frac{17}{4032}\psi^2+\mathcal{O}(\psi^{3})
\nonumber \\
B_2 &=& \frac{1}{25}+\frac{7}{600}\psi+\frac{1027}{100800}\psi^2
+\mathcal{O}(\psi^{3}) \nonumber
\\
B_3 &=&
\frac{1}{125}+\frac{43}{3000}\psi+\frac{1633}{72000}\psi^2+\mathcal{O}(\psi^{3}).
\nonumber
\end{eqnarray}

Using this information and integrating (\ref{Pg01}) we obtain the
genus one free energy $F^{(1)}_{\textrm{A-orbf.}}=
-\frac{s^5}{9}+\ldots$ The regularity of $F^{(1)}$, i.e. the
absence of log terms, is expected as there are no massless BPS
states at the Gepner-point. Because of this, the considerations in
Sec.\ref{boundaryconditions} imply also that the higher genus
amplitudes
\begin{eqnarray} \label{fgorb}
F^{(g)}_{\textrm{A-orbf.}} &=& \lim_{\bar{\tilde t}\rightarrow 0}
(\hat \omega_0)^{2(g-1)}(\frac{1-\psi}{5\psi})^{g-1}P_g
\end{eqnarray}
have no singularity at the orbifold point $\psi\sim 0$. This is in
accordance with the calculations in \cite{Katz:1999} and implies
that $\frac{P_g}{\psi^{\frac{3}{5}(g-1)}}$ is regular at $\psi\sim
0$.

The situation for the higher genus amplitudes  for the compact
${\cal O}(5)$ constraint in $\mathbb{P}^4$ is considerably
different from the one for the resolution ${\cal O}(-5)\rightarrow
\mathbb{P}^4$ of the $\mathbb{C}^5/\mathbb{Z}_5$. In normal
Gromov-Witten theory for genus $g>0$ on ${\cal O}(5)$ in
$\mathbb{P}^4$ there is no bundle whose Euler class of its pullback
from the ambient space $\mathbb{P}^4$ to the moduli space of
maps gives rise to a suitable measure on ${\cal M}_{g,\beta}$ that
counts the maps to the quintic. This is the same difficulties one
has to face for higher genus calculation for the orbifold GW
theory in ${\cal O}(5)$ in  $\mathbb{P}^4$, and it is notably
different\footnote{Since the brane bound state cohomology at
infinity can only be understood upon including higher genus
information, see Sec. \ref{symplecticinvariants} and
\ref{quinticdbrane}, the claims that one can learn essential
properties about the D-branes of the quintic at small volume from
the $\mathbb{C}^5/\mathbb{Z}_5$ orbifold might be overly
optimistic.} from the equivariant GW theory on
$\mathbb{C}^5/\mathbb{Z}_5$.

However we claim that our $F^{(g)}_{\textrm{A-orbf.}}$ predictions
from the $B$-model computation contain the information about the
light even RR states at the orbifold point in useful variables and
could in principle be checked  in the A-model by some version of
equivariant localisation. Below we give the first few order
results. They are available to genus $20$ at \cite{webpage}.
$$
\begin{array}{rl}
F^{(0)}_{\textrm{A-orbf.}}&=\frac{5\,s^3}{6} + \frac{5\,s^8}{1008} + \frac{5975\,s^{13}}{10378368} +
\frac{34521785\,s^{18}}{266765571072}+\ldots\\ [2 mm]
F^{(1)}_{\textrm{A-orbf.}}&= -\frac{s^5}{9} - \frac{163 \, s^{10}}{18144} - \frac{85031 \, s^{15}}{46702656} -
\frac{6909032915\, s^{20}}{20274183401472} +\ldots \\[2 mm]
F^{(2)}_{\textrm{A-orbf.}}&=\frac{155\,s^2}{18} - \frac{5\,s^7}{864} + \frac{585295\,s^{12}}{14370048} +
\frac{1710167735\,s^{17}}{177843714048}+\ldots\\[2 mm]
F^{(3)}_{\textrm{A-orbf.}}&=\frac{488305\,s^4}{9072} - \frac{3634345\,s^9}{979776} - \frac{1612981445\,s^{14}}{7846046208} -
  \frac{2426211933305\,s^{19}}{116115777662976} +\ldots\\[2 mm]
F^{(4)}_{\textrm{A-orbf.}}&=\frac{48550\,s}{567} + \frac{36705385\,s^6}{163296} + \frac{16986429665\,s^{11}}{603542016} +
  \frac{341329887875\,s^{16}}{70614415872}+\ldots\\[2 mm]
F^{(5)}_{\textrm{A-orbf.}}&=\frac{1237460905\,s^3}{224532} + \frac{108607458385\,s^8}{28740096} - \frac{2079654832074515\,s^{13}}{1553517149184} -
  \frac{50102421421803185\,s^{18}}{438808843984896}+\ldots\\ [2 mm]
\end{array}
$$

The holomorphic ambiguity (\ref{10-14-2.16}) is a power series
of $\psi$ starting from a constant term, so requiring
$\frac{P_g}{\psi^{\frac{3}{5}(g-1)}}$ to be regular imposes
\begin{equation}
\lceil\frac{3}{5}(g-1)\rceil
\end{equation}
number of relations in $a_i$ in (\ref{10-14-2.16}), where
$\lceil\frac{3}{5}(g-1)\rceil$ is the ceiling, i.e. the smallest
integer greater or equal to $\frac{3}{5}(g-1)$. We note that the
leading behaviour $\omega^{orb}_1\sim \psi^\frac{1}{d}$ with $d=5$
for the quintic, which is typical for an orbifold point in compact
Calabi-Yau, and which ``shields'' the singularity and diminishes
the boundary conditions at the orbifold point from $g-1$ to
$\lceil \frac{d-2}{d}(g-1)\rceil $. If this period is
non-vanishing at $\psi=0$ and it is indeed simply a constant for
all local cases~\cite{Huang2}, one gets $g-1$ conditions, which
together with the gap condition at the conifold and the constant
map information is already sufficient to completely solve the
model. An example of this type is ${\cal O}(-3)\rightarrow
\mathbb{P}^2$.

\subsection{Expansions around the conifold point $\psi=1$}
\label{quinticconifold}
An new feature of the conifold region is that there is an 
choice in picking the polarization, but as we will show 
the gap property is independent of this choice.       

A basis of solutions of the Picard-Fuchs equation around the
conifold point $\psi-1=\delta\sim 0$ is the following
\begin{equation}
\vec \Pi_c=\left(
\begin{array}{c}
\omega_0^c\\
\omega_1^c\\
\omega_2^c\\
\omega_3^c\\
\end{array}
\right)
=\left(
\begin{array}{l}
1+\frac{2\delta^3}{625}-\frac{83\delta^4}{18750}+\frac{757\delta^5}{156250}+\mathcal{O}(\delta^6)\\
\delta-\frac{3\delta^2}{10}+\frac{11\delta^3}{25}-\frac{217\delta^4}{2500}+\frac{889\delta^5}{15625}+\mathcal{O}(\delta^6)\\
\delta^2-\frac{23\delta^3}{30}+\frac{1049\delta^4}{1800}-\frac{34343\delta^5}{75000}+\mathcal{O}(\delta^6) \\
\omega_1^c\log(\delta)-\frac{9\,d^2}{20} - \frac{169\,d^3}{450} + 
  \frac{27007\,d^4}{90000} - 
  \frac{152517\,d^5}{625000}+\mathcal{O}(\delta^6)
\end{array}
\right)
\end{equation}
Here we use the superscript ``c'' in the periods to denote them as
solutions around the conifold point. We see that one of the
solutions $\omega_1^c$ is singled out as it multiplies the log in
the solution $\omega_3^c$. By a Lefshetz argument \cite{AGV} it
corresponds to the integral over the vanishing $S^3$ cycle $B_1$
and moreover a solution containing the log is the integral over
dual cycle $A_1$. Comparing with
(\ref{periods},\ref{dbranecharge}) shows in the Type IIA
interpretation that the $D6$ brane becomes massless. To determine
the symplectic basis we analytically continue the solutions
(\ref{periods}) from $\psi=\infty$ and get
\begin{equation}
\left(
\begin{array}{c}
F_0\\[2 mm]
F_1\\[2 mm]
X_0\\[2 mm]
X_1\\[2 mm]
\end{array}
\right)=
\left(\begin{array}{rrrr}
0&{\sqrt{5}\over 2 \pi i}&0&0\\[1 mm]
a-{11 i\over 2} g& b-{11i \over 2}h& c-{11i \over 2}r&0\\[1 mm]
d&e&f&-{ \sqrt{5} \over (2 \pi i)^2}\\
ig&ih& ir& 0\end{array}
\right)\left(
\begin{array}{c}
\omega_0^c\\[2 mm]
\omega_1^c\\[2 mm]
\omega_2^c\\[2 mm]
\omega_3^c\\[2 mm]
\end{array}
\right)
\end{equation}
Six of the real numbers
$a,\ldots,r$ are only known numerically\footnote{ 
$a= 6.19501627714957\ldots$,
$b= 1.016604716702582\ldots$, 
$c= -0.140889979448831\ldots$, 
$d=1.07072586843016\ldots$, 
$e= -0.0247076138044847\ldots$, 
$g= 1.29357398450411\ldots$, 
$h=\frac{ 2\,b\,g\,\pi -\left( {\sqrt{5}}\,d \right) }{2\,a\,\pi }$, 
$r=\frac{5 + 16\,c\,g\,{\pi }^3}{16\,a\,{\pi }^3}$,
$f=\frac{{\sqrt{5}}\,b + 8\,c\,d\,{\pi }^2}{8\,a\,{\pi }^2}$.}. 
Nevertheless we can give the symplectic form exactly in the new basis 
\begin{equation} 
\dd F_k\wedge \dd X_k = -\frac{1}{(2 \pi i)^3} \left(\frac{5}{2} \dd \omega_2^c\wedge\dd \omega_0^c +  
(-5)\dd \omega_3^c\wedge\dd \omega_1^c\right)\ .
\label{symplecticconifold} 
\end{equation}

The mirror map should be invariant under the conifold monodromy
and vanishing at the conifold. The vanishing period has $D6$
brane charge and is singled out to appear in the numerator of the
mirror map. The numerator is not fixed up to the fact that 
$\omega_3^c$ should not appear. The simplest mirror map compatible with 
symplectic form (\ref{symplecticconifold}) is
\begin{eqnarray} \label{dualmirrormap}
t_D(\delta)&:=&\frac{\omega_1^c}{\omega_0^c}=\delta-\frac{3
\delta^2}{10}+\frac{11\delta^3}{75}-
\frac{9\delta^4}{100}+\frac{5839 t_D^5}{93750}
+\mathcal{O}(t_D^6)\\
\delta(t_D)&=&t_D+\frac{3 t_D^2}{10}+\frac{t_D^3}{30}+\frac{t_D^4}{200}+\frac{169 t_D^5}{375000}
+\mathcal{O}(t_D^6)
\end{eqnarray}
We call this the dual mirror map and denote it $t_D$ to
distinguish from the large complex structure modulus case.

In the holomorphic limit $\bar{\delta}\rightarrow 0$, the Kahler
potential and metric should behave as $e^{-K}\sim \omega_0^c$ and
$G_{\delta\bar{\delta}}\sim \partial_\delta t_D$. We can find the
asymptotic behavior of various generators,
\begin{eqnarray}
X&=& \frac{1}{1-\psi}=-\frac{1}{\delta}
\nonumber \\
A&=&
-\frac{3}{5}-\frac{2}{25}\delta+\frac{2}{125}\delta^2-\frac{52}{9375}\delta^3+\mathcal{O}(\delta^4)
\nonumber \\
B&=&
\frac{6}{625}\delta^2-\frac{76}{9375}\delta^3+\frac{611}{93750}\delta^4
+\mathcal{O}(\delta^5)
\nonumber \\
B_2 &=&
\frac{12}{625}\delta-\frac{16}{3125}\delta^2+\frac{82}{46875}\delta^3
+\mathcal{O}(\delta^4)
\nonumber \\
B_3 &=&
\frac{12}{625}+\frac{28}{3125}\delta-\frac{78}{15625}\delta^2
+\mathcal{O}(\delta^3) \label{leadingconifold.}
\end{eqnarray}
Now we can expand
\begin{eqnarray} \label{10-14-2.21}
F^{(g)}_{\textrm{conifold}}=\lim_{\bar{\delta}\rightarrow
0}(\omega_0^c)^{2(g-1)}(\frac{1-\psi}{\psi})^{g-1}P_g
\end{eqnarray}
around the conifold point in terms of $t_D$ using the dual mirror
map (\ref{dualmirrormap}). 

Remarkably it turns out that  
shifts $\omega_0^c\rightarrow \omega_0^c+b_1\omega_1^c+b_2\omega_2^c$ 
does not affect the structure we are interested in. The fact that the 
$b_1$ shifts do not affect the amplitudes is reminiscent of the ${\rm
SL}(2,\mathbb{C})$ orbit theorem~\cite{Schmidt,CKS} and is proven
in App. \ref{sectioninvariance}. It is therefore reasonable to 
state the results in the more general polarization and define $\hat \omega^c_0= \omega^c_0 
+ b_1 \omega^c_1 + b_2 \omega^c_2$. We first determine the genus
$0$ prepotential checking consistency of the solutions with special 
geometry. Defining $\hat t_D=\frac{\omega_1^c}{\hat \omega_0^c}$ and 
$\Pi_{con}=(\hat \omega^c_0,\omega^c_1,\frac{5}{2}\omega^c_2,-5\omega^c_3)^T$ we get
\begin{equation}
\hat \Pi^{orb}=\hat \omega_0 \left(\begin{array}{c}
1\cr
\hat t_D\cr
2 F_{\textrm{conif.}}^{(0)}- \hat t_D \partial_{\hat t_D} 
F_{\textrm{conif.}}^{(0)}\cr
\partial_{\hat t_D} F_{\textrm{conif.}}^{(0)}
\end{array}
\right)\ .
\end{equation}
We can now calculate the $F^{(g)}$ in the generalized polarization. Since the normalization of
$\hat \omega_0^c$ is not fixed by the Picard-Fuchs equation, we 
pick the  normalization $b_0=1$ that will be convenient later on\footnote{The $b_0$ dependence can be restored noting
that each order in $\hat t_D$ is homogeneous in $b_i$.} Using the known expression for the ambiguity 
at genus $2$, $3$ \cite{BCOV, Katz:1999,Yamaguchi} for $g=0-3$ and Castelnuovos bound for $g>3$ we 
find the same interesting structure first observed in \cite{Huang}
\begin{equation} 
\begin{array}{rl} 
F_{\textrm{conif.}}^{(0)}&=-\frac{5}{2} \log (\hat t_D) \hat t_D^2 +\frac{5}{12}\left( 1 - 6b_1 \right)\hat t^3_D + 
  \left(\frac{5}{12} \left( b_1 - 3 b_2 \right)  -\frac{89}{1440}  - 
\frac{5}{4}\,b_1^2\right) \hat t_D^4 + {\cal O}(\hat t_D^5)\\ [2 mm]
F_{\textrm{conif.}}^{(1)}&=-\frac{\log (\hat t_D)}{12} + \left( \frac{233}{120} - \frac{113\,b_1}{12} \right) \,\hat t_D + 
  \left(\frac{233\,b_1}{120} - \frac{113\,{b_1}^2}{24} - 
     \frac{107 b_2}{12} -\frac{2681}{7200}\right) \,{\hat t_D}^2 + {{\cal O}(\hat t_D^3)}\\[2 mm]
F_{\textrm{conif.}}^{(2)}&=\frac{1}{240 {\hat t_D}^2} - \left(\frac{120373}{72000}   + \frac{11413 b_2}{144}\right) + 
  \left( \frac{107369}{150000} - \frac{120373\,b_1}{36000} + \frac{23533\,b_2}{720} - 
     \frac{11413 b_1 b_2}{72} \right) \,\hat t_D + {{\cal O}(\hat t_D^2)}\\ [2 mm] 
F_{\textrm{conif.}}^{(3)}&=\frac{1}{1008\,{\hat t_D}^4} - \left(
   \frac{178778753}{324000000}   + \frac{2287087\,b_2}{43200} + 
    \frac{1084235\,{b_2}^2}{864}\right) + {\cal O}(\hat t_D)\\[2 mm]
F_{\textrm{conif.}}^{(4)}&=\frac{1}{1440\,{\hat t_D}^6} - \left( \frac{977520873701}{3402000000000}   
+ \frac{162178069379\,b_2}{3888000000} + 
    \frac{5170381469\,{b_2}^2}{2592000} + \frac{490222589\,{b_2}^3}{15552}\right) + 
  {\cal O}(\hat t_D) \cr.
\end{array}
\end{equation}
As explained in sec. \ref{boundaryconditions} we expect this gap structure
to be present at higher genus as in (\ref{thegap}), i.e.
\begin{eqnarray}
F^{(g)}_{\textrm{conifold}}&=&\frac{(-1)^{g-1}B_{2g}}{2g(2g-2)\hat t_D^{2g-2}}+\mathcal{O}(\hat t_D^0).
\end{eqnarray}
If this is true then it will impose $2g-2$ conditions on the
holomorphic ambiguity (\ref{10-14-2.16}). We can further note that
because the prefactor in (\ref{10-14-2.21}) goes like
$\delta^{g-1}$, and the generator $X$ goes like $X\sim
\frac{1}{\delta}$, the terms $a_iX^i$ in holomorphic ambiguity
(\ref{10-14-2.16}) with $i\leq g-1$ do not affect the gap
structure in (\ref{thegap}). Therefore the gap structure fixes the
coefficients $a_i$ of $i=g, \cdots, 3g-3 $ in (\ref{10-14-2.16}),
but not the coefficients $a_i$ with $i\leq g-1$. 
Note that the choice of $b_i$, $i=0,1,2$ does not  affect the gap structure 
at all.  In App. \ref{sectioninvariance} it is proven that the generators 
of the modular forms in $P_g$ do not change under the shift $b_1$. 
The effect of this shift is hence merely a K\"ahler gauge 
transformation.   On the other hand we note that imposing the 
invariance of the gap structure under the $b_2$ shift does not 
give further conditions on the $a_i$, but it does affect 
the subleading expansion.

\subsection{Fixing the holomorphic ambiguity: a summary of results}

Putting all the information together, let us do a counting of
number of unknown coefficients. Originally we have $3g-2$
coefficients in (\ref{10-14-2.16}). The constant map calculation
\cite{Marino:1998, GVI, FP} in A-model large complex structure
modulus limit $\psi\sim \infty$ fixes one constant $a_0$. The
conifold expansion around $\psi\sim 1$ fixes $2g-2$ coefficients
$a_i$ of $i=g, \cdots, 3g-3$, and the orbifold expansion around
$\psi\sim 0$ further fixes $\lceil\frac{3(g-1)}{5}\rceil$
coefficients. So the number of unknown coefficients at genus $g$
is
\begin{eqnarray}
3g-2-(1+2g-2+\lceil\frac{3(g-1)}{5}\rceil)=[\frac{2(g-1)}{5}].
\label{boundarycountquintic}
\end{eqnarray}
This number is zero for genus $g=2,3$. So we could have computed
the $g=2,3$ topological strings using this information, although
the answers were already known. At $g\geq 4$ there are still
some unknown constants. However, in the A-model expansion when we
rewrite the Gromov-Witten invariants in terms of Gopakumar-Vafa
invariants, one can in principle use the Castelnuovos bound to fix
the $F^{(g)}$ up to genus 51. This will be shown in sect.
\ref{quinticdbrane}.

\section{One-parameter Calabi-Yau spaces with three regular singular points}
We  generalize the analysis for the quintic to other one K\"ahler
parameter Calabi-Yau three-folds, whose mirror $W$ has a
Picard-Fuchs-system with three regular singular points. Note that
this type of CY has been completely classified~\cite{DoranMorgan}
starting from the Riemann-Hilbert reconstruction of the
Picard-Fuchs equations given the monodromies and imposing the
special geometry property as well as integrality conditions on the
solutions of the latter. There are 13 cases whose mirrors can be
realized as hypersurfaces and complete intersections in weighted
projective spaces with trivial fundamental group\footnote{We also 
have some results on the cases with non-trivial fundamental group. 
They are available on request.}, and are well-known
in the mirror symmetry literature. A fourteenth case is
related to a degeneration of a two parameter model as pointed out
in~\cite{KKRS}. Five are obtained as a free discrete orbifold of
the former CY.

We focus on the $13$ former cases. In the notation of
\cite{Katz:1999} these complete intersections of degree
$(d_1,d_2,\cdots,d_k)$ in weighted projective spaces
$\mathbb{P}^n(w_1,\dots,w_l)$, Calabi-Yau manifolds are
abbreviated as $X_{d_1,d_2,\cdots,d_k}(w_1,\dots,w_l)$. For
example, the familiar quintic manifold is denoted as $X_{5}(1^5)$.
The list of 13 such examples is the following:
\begin{eqnarray}
&&X_{5}(1^5):
\vec{a}=(\frac{1}{5},\frac{2}{5},\frac{3}{5},\frac{4}{5}),~~
X_{6}(1^4,2):
\vec{a}=(\frac{1}{6},\frac{2}{6},\frac{4}{6},\frac{5}{6}),~~
 X_{8}(1^4,4):
\vec{a}=(\frac{1}{8},\frac{3}{8},\frac{5}{8},\frac{7}{8}),\nonumber \\
&& X_{10}(1^3,2,5):
\vec{a}=(\frac{1}{10},\frac{3}{10},\frac{7}{10},\frac{9}{10}),~~
X_{3,3}(1^6):
\vec{a}=(\frac{1}{3},\frac{1}{3},\frac{2}{3},\frac{2}{3}),\nonumber \\
&& X_{4,2}(1^6):
\vec{a}=(\frac{1}{4},\frac{1}{2},\frac{1}{2},\frac{3}{4}),
~~X_{3,2,2}(1^7):
\vec{a}=(\frac{1}{3},\frac{1}{2},\frac{1}{2},\frac{2}{3}),\nonumber \\
&& X_{2,2,2,2}(1^8):
\vec{a}=(\frac{1}{2},\frac{1}{2},\frac{1}{2},\frac{1}{2}),~~
X_{4,3}(1^5,2):
\vec{a}=(\frac{1}{4},\frac{1}{3},\frac{2}{3},\frac{3}{4}),
\nonumber \\
&&X_{4,4}(1^4,2^2):
\vec{a}=(\frac{1}{4},\frac{1}{4},\frac{3}{4},\frac{3}{4}),~~
X_{6,2}(1^5,3):
\vec{a}=(\frac{1}{6},\frac{1}{2},\frac{1}{2},\frac{5}{6}),
\nonumber \\
&&X_{6,4}(1^3,2^2,3):
\vec{a}=(\frac{1}{6},\frac{1}{4},\frac{3}{4},\frac{5}{6}),~~
X_{6,6}(1^2,2^2,3^2):
\vec{a}=(\frac{1}{6},\frac{1}{6},\frac{5}{6},\frac{5}{6}).
\nonumber
\end{eqnarray}
The examples satisfy the Calabi-Yau condition $\sum_i d_i=\sum_i
w_i$ required by the vanishing of the first Chern class. The
components of the vector $\vec{a}$ specify the Picard-Fuchs
operators for the mirror manifolds with $h^{2,1}=1$,
\begin{eqnarray} \label{PF-10-28-mh}
\{(\psi\frac{\partial}{\partial
\psi})^4-\psi^{-1}\prod_{i=1}^4(\psi\frac{\partial}{\partial
\psi}-a_i)\}\Pi=0.
\end{eqnarray}
The indices of the Picard-Fuchs equation satisfy $\sum_{i=1}^4
a_i=2$ and we have arranged $a_i$ in increasing order for later
convenience.

There are three singular points in the moduli space. The maximally
unipotent point is the large complex structure modulus limit
$\psi\sim \infty$ that has three logarithmic solutions for the
Picard-Fuchs equation. The conifold point is $\psi=1$ with three
power series solutions and one logarithmic solution for the
Picard-Fuchs equation. If the indices $a_i$ are not degenerate,
the Picard-Fuchs equation around the orbifold point $\psi=0$ has
four powers series solutions with the leading terms going like
$\psi^{a_i}$. Each degeneration of the indices generates a
logarithmic solution for the Picard-Fuchs equation around the
orbifold point $\psi=0$.
\subsection{The integration of the anomaly equation}
We can straightforwardly generalize the formalism in
\cite{Yamaguchi} to the above class of models.  The mirror map is
normalized as $t= \log(\frac{\prod_i d_i^{d_i}}{\prod_i
w_i^{w_i}}\psi)+\mathcal{O}(\frac{1}{\psi})$, so that the
classical intersection number in the prepotential is $F^{(0)}=
\frac{\kappa}{6}t^3+\cdots$ where $\kappa=\frac{\prod_i
d_i}{\prod_i w_i}$. The generators of the topological string
amplitudes are defined accordingly,
\begin{eqnarray}
&&A_p:=\frac{(\psi\partial\psi)^p
G_{\psi\bar{\psi}}}{G_{\psi\bar{\psi}}},~~B_p:=\frac{(\psi\partial\psi)^p
e^{-K}}{e^{-K}}, ~~(p=1,2,3,\cdots) \nonumber \\ &&
C:=C_{\psi\psi\psi}\psi^3,~~X:=\frac{1}{1-\psi}.
\end{eqnarray}
Here the familiar three point Yukawa coupling is
$C_{\psi\psi\psi}\sim \frac{\psi^{-2}}{1-\psi}$, and as in the
case of the quintic we denote $A:=A_1$ and $B:=B_1$. The generators
satisfy the derivative relations
\begin{eqnarray}
\psi\partial_\psi A_p=A_{p+1}-AA_p, ~~~ \psi\partial_\psi
B_p=B_{p+1}-BB_p,~~~ \psi\partial_\psi X=X(X-1), \nonumber
\end{eqnarray}
and the recursion relations
\begin{eqnarray}
B_4 &=&
(\sum_ia_i)XB_3-(\sum_{i<j}a_ia_j)XB_2+(\sum_{i<j<k}a_ia_ja_k)XB-(\prod_i
a_i)X, \nonumber \\
A_2 &=& -4B_2-2AB-2B+2B^2-2A+2XB+XA-r_0X-1,
\end{eqnarray}
where the first equation can be derived from the Picard-Fuchs
equation (\ref{PF-10-28-mh}) and the second equation is derived
from the special geometry relation up to a holomorphic ambiguity
denoted by the constant $r_0$. One can fix the constant $r_0$ by
expanding the generators around any of the singular points in the
moduli space. For example, as we will explain, the asymptotic
behaviors of various generators around the orbifold point are
$A_p=(a_2-a_1-1)^p+\mathcal{O}(\psi)$,
$B_p=a_1^p+\mathcal{O}(\psi)$, and $X=1+\mathcal{O}(\psi)$, so we
find the constant is
\begin{eqnarray} \label{constantr0-10-28}
r_0=a_1(1-a_1)+a_2(1-a_2)-1.
\end{eqnarray}
The polynomial topological amplitudes $P_g$ are defined by
$P^{(n)}_g:=C^{g-1}\psi^nC^{(g)}_{\psi^n}$, and satisfy the
recursion relations with initial data
\begin{eqnarray}
P^{(3)}_{g=0}&=&1, \nonumber \\
P^{(1)}_{g=1} &=&
(\frac{\chi}{24}-2)B-\frac{A}{2}+\frac{1}{12}(X-1)+\frac{s_1}{2},
\nonumber \\
P^{(n+1)}_g &=& \psi\partial_\psi
P^{(n)}_g-[n(A+1)+(2-2g)(B-\frac{X}{2})]P^{(n)}_g,
\end{eqnarray}
where $\chi$ is the Euler character of the Calabi-Yau space and
$s_1=2 c-\frac{5}{6}$ is a constant that can be fixed by the
second Chern class of the Calabi-Yau; see appendix A. We provide
the list of constants for the 13 cases of one-parameter Calabi-Yau
in the following table.

\begin{table}
\begin{centering}
\begin{tabular}{|r|r|r|r|r|r|r|r|}
\hline CY &$X_5(1^5)$ & $X_6(1^4,2)$ & $X_8(1^4,4)$ &
$X_{10}(1^3,2,5)$ & $X_{3,3}(1^6)$
\\ \hline
$\chi$ & $-200$ & $-204$ & $-296$ & $-288$ & $-144$ \\
\hline $s_1$ & $\frac{10}{3}$ & $\frac{8}{3}$ & $\frac{17}{6}$ &
$2$ & $\frac{11}{3}$ \\ \hline CY & $X_{4,2}(1^6)$ &
$X_{3,2,2}(1^7)$ &$X_{2,2,2,2}(1^8)$ & $X_{4,3}(1^5,2)$ &
$X_{4,4}(1^4,2^2)$  \\ \hline $\chi$ & $-176$ & $-144$  & $-128$ &
$-156$ & $-144$  \\ \hline $s_1$  & $\frac{23}{6}$ &
$\frac{25}{6}$ & $\frac{9}{2}$ & $\frac{19}{6}$ & $\frac{5}{2}$ \\
\hline CY & $X_{6,2}(1^5,3)$ & $X_{6,4}(1^3,2^2,3)$ &
$X_{6,6}(1^2,2^2,3^2)$ & &\\ \hline  $\chi$ & $-256$ & $-156$ &
$-120$  &
&\\ \hline  $s_1$ & $\frac{7}{2}$ & $\frac{11}{6}$ & $1$  & & \\
\hline
\end{tabular} \caption{Euler numbers and the constant $s_1$}
\end{centering}
\end{table}

After changing variables to a convenient basis from
$(A,B,B_2,B_3,X)$ to $(u,v_1,v_2,v_3,X)$ as the following:
\begin{eqnarray}
&&B=u,~~A=v_1-1-2u,~~B_2=v_2+uv_1, \nonumber \\ &&
B_3=v_3-uv_2+uv_1X-(r_0+1)u X. \nonumber
\end{eqnarray}
One can follow \cite{Yamaguchi} and use the BCOV formalism to show
that $P_g$ is a degree $3g-3$ inhomogeneous polynomial of
$v_1,v_2,v_3, X$ with the assigned degrees $1,2,3,1$ for $v_1,
v_2, v_3, X$ respectively. The BCOV holomorphic anomaly equation
becomes
\begin{eqnarray}
(\frac{\partial}{\partial{v_1}}-u\frac{\partial }{\partial
{v_2}}-u(u+X)\frac{\partial}{\partial{v_3}})P_g
=-\frac{1}{2}(P^{(2)}_{g-1}+\sum_{r=1}^{g-1}P^{(1)}_rP^{(1)}_{g-r}).
\end{eqnarray}
As in the case of the quintic, the holomorphic anomaly equation
determines the polynomial $P_g$ up to a holomorphic ambiguity
\begin{eqnarray}
f=\sum_{i=0}^{3g-3} a_iX^i.
\end{eqnarray}

\subsection{The boundary behaviour}
The constant term $a_0$ is fixed by the known constant map
contribution in the A-model expansion
\begin{eqnarray}
F^{g}_{\textrm{A-model}}=\lim _{t\rightarrow \infty}
\omega_0^{2(g-1)}(\frac{1-\psi}{\kappa \psi})^{g-1}P_g,
\end{eqnarray}
where $\omega_0$ is the power series solution in the large complex
structure modulus limit.

The main message here is that structure of the conifold expansion
is universal. For all cases of Calabi-Yau spaces, the Picard-Fuchs
equation around $z=\psi-1$ has four solutions that go like
$\omega_0= 1+\mathcal{O}(z)$, $\omega_1= z+\mathcal{O}(z^2)$,
$\omega_2=z^2+\mathcal{O}(z^3)$, and
$\omega_4=\omega_1\log(z)+\mathcal{O}(z^4)$. We define a dual
mirror map $t_D=\frac{\omega_1}{\omega_0}=z+\mathcal{O}(z^2)$,
expand the topological string amplitudes in terms of $t_D$, and
impose the following gap structure in the conifold expansion:
\begin{eqnarray}
F^{(g)}_{\textrm{conifold}} &=& \lim_{\bar{z}\rightarrow
0}\omega_0^{2(g-1)}(\frac{1-\psi}{\psi})^{g-1}P_g \nonumber \\
&=&
\frac{(-1)^{g-1}B_{2g}}{2g(2g-2)t_D^{2g-2}}+\mathcal{O}(t_D^0).
\end{eqnarray}
This fixes $2g-2$ coefficients in the holomorphic ambiguity.

On the other hand, we discover a rich variety of singularity
structures around the orbifold point $\psi=0$. A natural
symplectic basis of solutions of the Picard-Fuchs equation
(\ref{PF-10-28-mh}) is picked out by the fractional powers of the
leading terms. As in the case of quintic for our purpose we only
need the first two solutions $\omega_0$, $\omega_1$. The leading
behaviors are
\begin{enumerate}
\item If $a_2>a_1$, then we have
$\omega_0=\psi^{a_1}(1+\mathcal{O}(\psi))$,
$\omega_1=\psi^{a_2}(1+\mathcal{O}(\psi))$.
\item If $a_2=a_1$, then we have
$\omega_0=\psi^{a_1}(1+\mathcal{O}(\psi))$,
$\omega_1=\omega_0(\log(\psi)+\mathcal{O}(\psi))$.
\end{enumerate}
In both cases we can define a mirror map around the orbifold point
as $s=\frac{\omega_1}{\omega_0}$. Using the behaviors of
K\"ahler potential $e^{-K}\sim \omega_0$ and metric
$G_{\psi\bar{\psi}}\sim
\partial_\psi s$ in the
holomorphic limit, we can find the asymptotic expansion of various
generators. In both cases $a_2>a_1$ and $a_2=a_1$, the leading
behaviors are non-singular:
\begin{eqnarray}
A_p &=& (a_2-a_1-1)^p+\mathcal{O}(\psi) \nonumber \\
B_p &=& a_1^p+\mathcal{O}(\psi).
\end{eqnarray}
This can be used to fix a holomorphic ambiguity
(\ref{constantr0-10-28}) relating $A_2$ to other generators. On
the other hand, since the constant (\ref{constantr0-10-28})  can
be also derived at other singular points of the moduli space, this
also serves as a consistency check that we have chosen the correct
basis of solutions $\omega_0$, $\omega_1$ at the orbifold point.

The orbifold expansion of the topological string amplitudes are
\begin{eqnarray}
F^{(g)}_{\textrm{orbifold}}=\lim_{\bar{\psi}\rightarrow 0
}\omega_0^{2(g-1)}(\frac{1-\psi}{\psi})^{g-1}P_g\sim
\frac{P_g}{\psi^{(1-2a_1)(g-1)}}
\end{eqnarray}
We can expand the polynomial $P_g$ around the orbifold point
$\psi=0$. Generically, $P_g$ is power series of $\psi$ starting
from a constant term, so $F^{(g)}_{\textrm{orbifold}}\sim
\frac{1}{\psi^{(1-2a_1)(g-1)}}$. Since $\sum_ia_i=2$ and $a_1$ is
the smallest, we know $a_1\leq \frac{1}{2}$ and the topological
string amplitude around orbifold point is generically singular.
Interestingly, we find the singular behavior of the topological
strings around the orbifold point is not universal and falls into
four classes:

\begin{enumerate}
\item This class includes all 4 cases of hypersurfaces and 4 other cases of complete intersections.
They are the Calabi-Yau spaces $X_5(1^5)$, $X_6(1^4,2)$,
$X_8(1^4,4)$, $X_{10}(1^3,2,5)$, $X_{3,3}(1^6)$,
$X_{2,2,2,2}(1^8)$, $X_{4,4}(1^4,2^2)$, $X_{6,6}(1^2,2^2,3^2)$.
For these cases the singularity at the orbifold point is cancelled
by the series expansion of the polynomial $P_g$. The requirement
of cancellation of singularity in turn imposes
\begin{eqnarray}
\lceil(1-2a_1)(g-1)\rceil
\end{eqnarray}
conditions on the holomorphic ambiguity in $P_g$. Taking into
account the A-model constant map condition and the boundary
condition from conifold expansion, we find the number of unknown
coefficients at genus $g$ is
\begin{eqnarray} \label{number1-10-28}
[2a_1(g-1)].
\end{eqnarray}
Notice for the Calabi-Yau $X_{2,2,2,2}(1^8)$ the cancellation is
trivial since in this case $a_1=\frac{1}{2}$ and topological
strings around the orbifold point are generically non-singular, so
this does not impose any boundary conditions.

\item This class of Calabi-Yau spaces include $X_{4,2}(1^6)$ and
$X_{6,2}(1^5,3)$. We find the singularity at the orbifold point is
not cancelled by the polynomial $P_g$, but it has a gap structure
that closely resembles the conifold expansion. Namely, when we
expand the topological strings in terms of the orbifold mirror map
$s=\frac{\omega_1}{\omega_0}$, we find
\begin{eqnarray}
F^{(g)}_{\textrm{orbifold}}=\frac{C_g}{s^{2(g-1)}}+\mathcal{O}(s^0)
\end{eqnarray}
where in our normalization convention, the constant is
$C_g=\frac{B_{2g}}{2^{5g-4}(g-1)g}$ for the $X_{4,2}(1^6)$ model,
and $C_g=\frac{B_{2g}}{3^{3(g-1)}4(g-1)g}$ for the
$X_{6,2}(1^5,3)$ model. Since the mirror map goes like
$s\sim \psi^{a_2-a_1}$, and the $P_g$ is a power series of
$\psi$, this imposes
\begin{eqnarray}
\lceil 2(a_2-a_1)(g-1) \rceil
\end{eqnarray}
conditions on the holomorphic ambiguity in $P_g$. In both cases of
$X_{4,2}(1^6)$ and $X_{6,2}(1^5,3)$, we have $a_2=\frac{1}{2}$, so
the number of un-fixed coefficients is the same as the models in
the first class, namely
\begin{eqnarray}
[2a_1(g-1)].
\end{eqnarray}

\item This class of Calabi-Yau spaces include only the Calabi-Yau $X_{3,2,2}(1^7)$.
For this model, the singularity around the orbifold point does not
cancel, so there is no boundary condition imposed on the
holomorphic ambiguity in $P_g$. At genus $g$ we are simply left
with $g-1$ unknown coefficients.

\item This class of Calabi-Yau spaces include $X_{4,3}(1^5,2)$ and
$X_{6,4}(1^3,2^2,3)$. For this class of models, the singularity at
the orbifold point is partly cancelled. Specifically, we find
$\frac{P_g}{\psi^{(1-2a_1)(g-1)}}$ is not generically regular at
$\psi\sim 0$, but $\frac{P_g}{\psi^{(1-2a_2)(g-1)}}$ is always
regular. This then imposes
\begin{eqnarray}
\lceil(1-2a_2)(g-1)\rceil
\end{eqnarray}
conditions on holomorphic ambiguity in $P_g$, and the number of
unknown coefficients at genus $g$ is now
\begin{eqnarray}
[2a_2(g-1)].
\end{eqnarray}

\end{enumerate}

It looks like we have the worst scenarios in the two models
$X_{2,2,2,2}(1^8)$ and $X_{3,2,2}(1^7)$, where we essentially get
no obvious boundary conditions at the orbifold point $\psi=0$.
However after a closer examination, we find some patterns in the
leading coefficients of the orbifold expansion as the followings.
In our normalization convention, we find the leading constant
coefficients of $X_{2,2,2,2}(1^8)$ model is
\begin{eqnarray}
F^{(g)}_{\textrm{orbifold}}=\frac{(21\cdot2^{2g-2}-5)(-1)^{g}B_{2g}B_{2g-2}}{2^{2g-3}g(2g-2)(2g-2)!}
+\mathcal{O}(\psi)
\end{eqnarray}
whereas for $X_{3,2,2}(1^7)$ model the leading coefficients are
\begin{eqnarray} \label{twoparticles-12-01}
F^{(g)}_{\textrm{orbifold}}=
\frac{(7\cdot2^{2g-2}-1)B_{2g}}{2^{4g-3}3^{2g-2}(g-1)g}
\frac{1}{s^{2g-2}}+\mathcal{O}(\frac{1}{s^{2g-8}})
\end{eqnarray}
These leading coefficients provide one more boundary condition for
the models $X_{2,2,2,2}(1^8)$ and $X_{3,2,2}(1^7)$, although this
is not much significant at large genus. On the other hand, we
observe the leading coefficients of $X_{2,2,2,2}(1^8)$ model are
similar to the constant map contribution of Gromov-Witten
invariants except the factor of $(21\cdot2^{2g-2}-5)$, whereas the
leading coefficients of $X_{3,2,2}(1^7)$ model are similar to the
conifold expansion except the factor of $(7\cdot2^{2g-2}-1)$.
These non-trivial factors can not be simply removed by a different
normalization of variables and therefore contain useful
information. In fact in the latter case of $X_{3,2,2}(1^7)$ model,
the factor of $(7\cdot2^{2g-2}-1)$ will motivate our physical
explanations of the singularity in a moment.

As shown in \cite{Katz:1999}, see also sec.
\ref{symplecticinvariants}, for a fixed genus $g$, the
Gopakumar-Vafa invariants $n^g_d$ are only non-vanishing when the
degree $d$ is bigger than $a_g$, where $a_g$ is a model dependent
number with weak genus dependence, in particular for large $g$ one
has $d_{\textrm{min}}-1 = a_g \sim \sqrt{g}$. So long as the
number of zeros in the low degree Gopakumar-Vafa invariants are
bigger than the number of unknown coefficients that we determine
above using all available boundary conditions, we have a
redundancy of data to compute the topological strings recursively
genus by genus, and are able to make non-trivial checks of our
computations. For all of the 13 cases of one-parameter Calabi-Yau
spaces, we are able to push the computation to very high genus. So
far our calculations are limited only by the power of our
computational facilities.

We now propose a ``phenomenological'' theory of the singularity
structures at the point with rational branching. Our underlying philosophy is
that a singularity of $F^{(g)}$ in the moduli space can only be
generated if there are charged massless states near this point of
moduli space. This is already familiar from the behaviors at
infinity $\psi=\infty$ and conifold point $\psi=1$. At infinity
$\psi=\infty$ the relevant charged states are massive $D2-D0$
brane bound states and therefore $F^{(g)}$ are regular, whereas at
the conifold point there is a massless charged state from a $D3$
brane wrapping a vanishing 3-cycle, and this generates the gap
like singularity at the conifold point as we have explained. We
should now apply this philosophy to the much richer behavior at
the orbifold point $\psi=0$. We discuss the four classes of models
in the same order as mentioned above.
\begin{enumerate}
\item We argue for this class of models the $F^{(g)}$ are regular
at the orbifold point because there is no massless charged state.
A necessary condition would be that the mirror map parameter
$s$ is non-zero at the orbifold point, since as we have
learned there are D-branes wrapping cycles whose charge and mass
are measured by $s$. This is clear for the complete
intersection cases, namely $X_{3,3}(1^6)$, $X_{2,2,2,2}(1^8)$,
$X_{4,4}(1^4,2^2)$, $X_{6,6}(1^2,2^2,3^2)$, because for these
models the first two indices of the Picard-Fuchs equation is
degenerate $a_2=a_1$, therefore the mirror map goes like
\begin{eqnarray}
s\sim \log{\psi}\rightarrow \infty
\end{eqnarray}
and we see that the D-branes are very massive and generate
exponentially small corrections just like the situation at
infinity $\psi=\infty$. As for the hypersurface cases $X_5(1^5)$,
$X_6(1^4,2)$, $X_8(1^4,4)$, $X_{10}(1^3,2,5)$, we comment that
although naively the mirror map goes like $s\sim
\psi^{a_2-a_1}\rightarrow 0 $, there is a change of basis under
which the generators are invariant as explained in Appendix
\ref{sectioninvariance}, and which could make the periods goes
like $\omega_0\sim\omega_1\sim \psi^{a_1}$, therefore the mirror
map becomes finite at the orbifold point. This is also consistent
with the basis at orbifold point we obtained by analytic
continuation from infinity $\psi=\infty$. In fact we checked that the regularity 
of $F^{(g)}$ is not affected when we take $s$ to be the ratio of generic 
arbitrary linear combinations of the periods $\omega_i$, $i=0,1,2,3$.  

\item We argue this class of models have a conifold like structure
because of the same mechanism we have seen for the conifold point
$\psi=1$. This is consistent with the fact that the degenerate
indices in these cases, e.g. models $X_{4,2}(1^6)$ and
$X_{6,2}(1^5,3)$, are the middle indices, namely we have
$a_2=a_3=\frac{1}{2}$. Therefore the Picard-Fuchs equation
constrains one of periods to be a power series proportional to
\begin{eqnarray}
\omega_1\sim \psi^{\frac{1}{2}}
\end{eqnarray}
Since there is another period that goes like $\omega_0\sim
\psi^{a_1}$, the mirror map goes like $s\sim
\psi^{\frac{1}{2}-a_1}\rightarrow 0$ and it is not possible to
change the basis in a way such that the mirror map is finite.
Integrating out a charged nearly massless particle generates the
gap like conifold singularity as we have explained.

\item For the model $X_{3,2,2}(1^7)$ we argue there are two
massless states near the orbifold point. Since the middle indices
also degenerate $a_2=a_3=\frac{1}{2}$, we can apply the same
reasoning from the previous case and infer that the mirror map
goes like $s\sim \psi^{\frac{1}{6}}$ and there must be at
least one charged massless state from a D3 brane wrapping
vanishing 3-cycle. However the situation is now more complicated.
We do not find a gap structure in the expansion of
$F^{(g)}_{\textrm{orbifold}}$, and the leading coefficients differ
from the usual conifold expansion by a factor of $(7\cdot
2^{2g-2}-1)$ as observed in (\ref{twoparticles-12-01}). These can
be explained by postulating that there are two massless particles
in this case whose masses are $m$ and $2m$. This fits nicely with
the $2^{2g-2}$ power in the intriguing factor of $(7\cdot
2^{2g-2}-1)$, and also explains the absence of gap structure by
the possible interactions between the two massless particles.

\item Finally, we discuss the cases of models $X_{4,3}(1^5,2)$ and
$X_{6,4}(1^3,2^2,3)$. The indices $a_i$ are not degenerate in
these cases. What makes these models different from the
hypersurface cases is the fact that the ratio $\frac{a_2}{a_1}$ is
now not an integer in these cases. This makes it difficult to
change the basis such that the mirror map is finite. We conjecture
that the mirror map indeed goes like $s\sim \psi^{a_2-a_1}$ and
therefore there exist massless particle(s) whose masses are
proportional to $s$ and who are responsible for generating the
singularity of $F^{(g)}$ at the orbifold point. This is consistent
with our analytic continuation analysis in Appendix
\ref{appendixD-12-06}. We note that a necessary non-trivial
consequence of this scenario would be that the
$F^{(g)}_{\textrm{orbifold}}$ is no more singular than
$\frac{1}{s^{2g-2}}$, i.e. the product
\begin{eqnarray}
s^{2g-2}F^{(g)}_{\textrm{orbifold}}\sim
\frac{P_g}{\psi^{(1-2a_2)(g-1)}}
\end{eqnarray}
should be regular. This is precisely what we observe
experimentally.

\end{enumerate}

\section{Symplectic invariants at large radius}
\label{symplecticinvariants} The coefficients of the large radius
expansion of the $\cF^{(g)}=\lim_{{\bar t}\rightarrow \infty}
F^{(g)}(t,\bar t)$  have an intriguing conjectural interpretation
as symplectic invariants of $M$. First of all we have
\begin{equation}
\cF^{(g)}(q)=\sum_{\beta} r_\beta^{(g)} q^{\beta}\ ,
\end{equation}
where $r^{(g)}_\beta\in \mathbb{Q}$ are the {\sl Gromow-Witten
invariants} of holomorphic maps. Secondly the {\sl Gopakumar-Vafa
invariants}~\cite{GVII} count the cohomology of the $D_0-D_2$
bound state moduli space, see also~\cite{Katz:1999},  and are
related to the
\begin{equation}
\begin{array}{rcl}
\cF(\lambda,t)&=&\ds{\sum_{g=0}^\infty \lambda^{2 g-2} \cF^{(g)}}(t)\\
&=&\ds{{c(t)\over \lambda^2}+l(t)+\sum_{g=0}^\infty \sum_{\beta\in H_2(M,\ZZ)}\sum_{m=1}^\infty n_{\beta}^{(g)} {1\over m}
\left(2 \sin {m \lambda \over 2}\right)^{2 g-2}  q^{\beta m}}\ .
\end{array}
\label{schwingerloopd2d0}
\end{equation}
Here $c(t)$ and $l(t)$ are some cubic and linear polynomials in $t$, which follow
from the leading behaviour of $\cF^{(0)}$ and $\cF^{(1)}$  as explained in (\ref{prepotential})
and (\ref{F1}).
With $q_\lambda=e^{i\lambda}$ we can write a product form\footnote{Here we dropped the $\exp({c(t)\over \lambda^2}+l(t))$
factor of the classical terms  at genus $0,1$.} for the partition function $Z^{\rm hol}=\exp(\cF^{\rm hol})$
\begin{equation}
Z^{\rm hol}_{\rm GV}(M,\lambda,q)=\prod_{\beta}\left[
\left(\prod_{r=1}^\infty (1-q_\lambda^r q^\beta)^{r n_\beta^{(0)}}\right)
\prod_{g=1}^\infty \prod_{l=0}^{2g-2}(1-q_\lambda^{ g-l-1} q^\beta)^{(-1)^{g+r}
\left(2 g-2\atop l\right) n_\beta^{(g)}}\right]\
\label{zhol}
\end{equation}
in terms of the {\sl Gopakumar-Vafa invariants} $n_\beta^{(g)}$. Based on the partition
functions there is a conjectural relation of the latter to the {\sl Donaldson-Thomas invariants}
$\tilde n_\beta^{(g)}$, which are invariants of the moduli space $I_k(M,\beta)$ of
ideal sheaves $\cI$ on $M$.  Defining $Z^{\rm hol}_{\rm DT}(M,q_{\lambda},q)=
\sum_{\beta,k\in {\ZZ}} {\tilde n}^{(k)}_{\beta} q_\lambda^k q^\beta$ one expects~\cite{DTGW}
\begin{equation}
Z^{\rm hol}_{\rm GV}(M,q_{\lambda},q)M(q_\lambda)^{\frac{\chi(M)}{2}}=Z^{\rm hol}_{\rm DT}(M,-q_{\lambda},q)\ ,
\label{eq:GVDT}
\end{equation}
where the McMahon function  is defined as $M(q_\lambda):=\prod_{n\ge 0} {1\over (1-q_\lambda^n)^n}$.
We will give below the information about the ${\cal F}^{(g)}$ in terms of the
{\sl Gopakumar-Vafa invariants} and give a  more detailed account of the data
of  the symplectic invariants on the webpage \cite{webpage}.

\begin{table}
\begin{centering}
\begin{tabular}{|r|rrrrrr|}
\hline
g &d=1 &d=2 &d=3 &d=4 &d=5 &d=6  \\
\hline
\, 0& 2875& 609250& 317206375& 242467530000& \!229305888887625&\! 248249742118022000 \\
1& 0    &     0&    609250&   3721431625&  12129909700200&  31147299733286500 \\
2& 0    &     0&         0&       534750&     75478987900&    871708139638250 \\
3& 0    &     0&         0&         8625&       -15663750&     3156446162875  \\
4& 0    &     0&         0&            0&           49250&       -7529331750 \\
5& 0    &     0&         0&            0&            1100&         -3079125 \\
6& 0    &     0&         0&            0&              10&            -34500 \\
7& 0    &     0&         0&            0&               0&                0  \\
\hline
\end{tabular}
\vskip 2 pt
\begin{tabular}{|r|rrr|}
\hline
g &d=7 & d=8 &d=9 \\
\hline
0& \!\!\! 295091050570845659250& \!\!\!\!\!\! 375632160937476603550000& \!\!\!\!\!\! 503840510416985243645106250\\
1& \!\!\! 71578406022880761750& \!\!\!\!\!\! 154990541752961568418125& \!\!\!\!\!\! 324064464310279585657008750 \\
2& 5185462556617269625& 22516841063105917766750& 81464921786839566502560125 \\
3& 111468926053022750& 1303464598408583455000& 9523213659169217568991500 \\
4& 245477430615250& 25517502254834226750& 507723496514433561498250 \\
5& -1917984531500& 46569889619570625& 10280743594493108319750 \\
6& 1300955250& -471852100909500& 30884164195870217250 \\
7& 4874000& 2876330661125&-135197508177440750 \\
8& 0& -1670397000& 1937652290971125 \\
9& 0& -6092500& -12735865055000\\
10& 0& 0& 18763368375\\
11& 0& 0& 5502750\\
12& 0& 0& 60375\\
13& 0& 0&  0\\
\hline
\end{tabular}
\begin{tabular}{|r|rr|}
\hline
g &d=10 & d=11 \\
\hline
0& \quad \,\,\,\,704288164978454686113488249750&  \quad \,\,\,\,\,   1017913203569692432490203659468875\\
1& 662863774391414096742406576300& 1336442091735463067608016312923750 \\
2& 261910639528673259095545137450& 775720627148503750199049691449750 \\
3& 52939966189791662442040406825 & 245749672908222069999611527634750 \\
4& 5646690223118638682929856600  &44847555720065830716840300475375\\
5& 302653046360802682731297875   & 4695086609484491386537177620000\\
6& 6948750094748611384962730     &267789764216841760168691381625\\
7& 40179519996158239076800       &7357099242952070238708870000\\
8& -25301032766083303150         &72742651599368002897701250\\
9& 1155593062739271425           &140965985795732693440000\\
10& -17976209529424700           &722850712031170092000 \\
11& 150444095741780              &-18998955257482171250\\
12& -454092663150                & 353650228902738500 \\
13&   50530375                   &-4041708780324500\\
14&  -286650                     &22562306494375\\
15&   -5700                      &-29938013250\\
16&    -50                       & -7357125\\
17&     0                        &  -86250\\
18&     0                        &      0\\
\hline
\end{tabular}
\caption{BPS invariants $n^g_d$ on the Quintic hypersurface in $\mathbb{P}^4$. See also Table 3.}
\end{centering}
\end{table}

\subsection{Castelnuovo's theory and the cohomology of the BPS state moduli space}

Let us give checks of the numbers using techniques of algebraic
geometry and the description of the BPS moduli space and its
cohomology developed in \cite{GVII,Katz:1999}. The aim is to check
the gap condition in various geometric settings, namely
hypersurfaces and complete intersections in (weighted) projective
spaces discussed before. According to \cite{GVII,Katz:1999} the
BPS number of a given charge, i.e. degree $d$, can be calculated
from cohomology of the moduli  space $\hat {\cal M}$ of a
$D_2-D_0$ brane system. The latter is the fibration of the
Jacobian $T^{2 \tilde g}$ of a genus ${\tilde g}$ curve over its
moduli space of deformations ${\cal M}$ . Curves of arithmetic
genus $g<{\tilde g}$ are degenerate curves, in the simplest case
with $\delta={\tilde g}-g$ nodes. Their BPS numbers are calculated
using the Euler numbers of relative Hilbert schemes ${\cal
C}^{(i)}$ of the universal curve (${\cal C}^{(0)}={\cal M}$,
${\cal C}^{(1)}$ is the universal curve, etc) in simple situations
as follows:
\begin{equation}
\begin{array}{rl}
n_d^{g}&={\displaystyle n_d^{{\tilde g} -\delta}=(-1)^{{\rm
dim}({\cal M})+\delta}\sum_{p=0}^\delta b( {\tilde g}-p,\delta-p)
e({\cal C}^{(p)})},
\\[ 3 mm]
b(g,k)&={2\over k!} (g-1)\prod_{i=1}^{k-1} (2 g - (k+2) +i),\qquad   b(g,0)=0.
\end{array}
\label{bpsformula}
\end{equation}

As explained in \cite{Katz:1999} curves in  projective spaces
meeting the  quintic are either plane curves in $\mathbb{P}^2$,
curves in $\mathbb{P}^3$, or $\mathbb{P}^4$. In all case one gets
from Castelnuovo theory a bound on $g$, which grows for large $d$
like  $g(d)\sim d^2$. For a detailed exposition of curves in
projective space see~\cite{Harris}. Using this information one can
determine which curves above is realized and contributes to the
BPS numbers. These statements generalize to the  hypersurfaces and
complete intersections with one K\"ahler modulus in weighted
projective spaces. In particular the qualitative feature $g(d)\sim
d^2$ of the bound for large $d$ carries over. We note for
later convenience that to go from a smooth curve of genus $\tilde
g$ to a curve with arithmetic genus $g=\tilde g-\delta$ by
enforcing $\delta$ nodes we get from (\ref{bpsformula})
\begin{equation}
\begin{array}{rl}
n^{\tilde g-1}_d&=(-1)^{{\rm dim}({\cal M})+1}\left( e({\cal C})+ (2\tilde g-2) e(\cal M)\right)\\
n^{\tilde g-2}_d&=(-1)^{{\rm dim}({\cal M})+1}\left(e({\cal C}^{(2)})+(2 \tilde g-4) e({\cal C})+{1\over 2}(2 \tilde g-2) (2 \tilde g-5)e(\cal M)\right)\ . \\
\end{array}
\label{BPSspecial}
\end{equation}

\subsection{D-branes on the quintic}
\label{quinticdbrane}
One consequence of our global understanding of the $F^{(g)}$
is that we can make detailed statements about the `number' of
$D$-brane states for the quintic at large radius.
 We focus on $d=5$, because
there is a small numerical flaw in the analysis of
\cite{Katz:1999}, while the right numerics confirms the gap
structure quite significantly. In this case the complete
intersection with multidegree $(1,1,5)$ is a plane curve with
genus $\tilde g=(d-1)(d-2)/2=6$, while the other possibilities
have at most genus $g=2$. Curves of $(g=\tilde g,d)=(6,5)$ are
therefore smooth plane curves with $\delta=0$ and according to
(\ref{bpsformula}) their BPS number is simply $n^g_d=(-1)^{{\rm
dim} \cal M} e({\cal M})$. Since $d=5$ their moduli space is
simply the moduli space of $\mathbb{P}^2$'s in $\mathbb{P}^4$,
${\cal M}$ is the Grassmannian\footnote{The space of
$\mathbb{P}^k$'s in  $\mathbb{P}^n$ , which we call
${\mathbb{G}}(k,n)$, is also the space of $k+1$ complex
dimensional subspaces in an $n+1$ dimensional complex vector
space, which is often alternatively denote as $G(k+1,n+1)$.}
${\mathbb{G}}(2,4)$. Grassmannians ${\mathbb{G}}(k,n)$ have
dimensions $(k+1)(n-k)$ and their Euler number can be calculated
most easily by counting toric fixed points to be
$\chi({\mathbb{G}}(k,n))=\left(n+1\atop k+1\right)$.  We get
$n^6_5=(-1)^6 10=10$.

For the $(g,d)=(5,5)$ curves we have to determine the Euler number
of the universal curve ${\cal C}$, which is a fibration $\pi:{\cal
C} \rightarrow {\cal M}$ over ${\cal M}$. To get  an geometric
model for ${\cal C}$ we consider the projection $\tilde \pi:{\cal
C}\rightarrow X$. The fiber over a point $p\in X$ is the  set of
$\IP^2$'s in $\IP^4$ which contain the point $p$. This is described
as the space of $\mathbb{P}^1$'s in $\mathbb{P}^3$ i.e.
${\mathbb{G}}(1,3)$ with~\footnote{${\mathbb{G}}(1,3)$ is Plucker
embedded in $\mathbb{P}^5$ as a quadric (degree 2). From the
adjunction formula we also  get $\chi({\mathbb{G}}(1,3))=6$.}
$\chi({\mathbb{G}}(1,3))=6$. As the fibration $\tilde \pi$ is
smooth we obtain $e({\cal C})=\chi(X)
\chi({\mathbb{G}}(1,3))=-200\cdot 6=-1200$. Applying  now
(\ref{BPSspecial}) we get $n^{5}_5=(-1)^5(-1200+ (2 \cdot 6-2)
10)=1100$.

The calculation of $n^4_4$ requires the  calculation  of  $e({\cal C}^{(2)})$. The model
for ${\cal C}^{(2)}$ is constructed from the fibration  $\hat \pi:{\cal C}^{(2)}\rightarrow {\rm Hilb}^2(X)$
as follows. A point in ${\rm Hilb}^2(X)$ are either two distinct points or one point of multiplicity $2$ with  distinct
tangent direction.  In both cases the
fiber over $P \in {\rm Hilb}^2(X)$ is an $\mathbb{P}^2$ passing though $2$ points in $\mathbb{P}^4$,
which is a  $\mathbb{P}^2$. The fibration is smooth and it remains to calculate the Euler number
of the basis.
There are nice product formulas for the Euler number of symmetric
products of surfaces modded out by $S_n$. For surfaces it is more
cumbersome. We calculate the Euler number $e({\rm
Sym}^2(X))=\left(-199\atop 2\right)$. ${\rm Hilb}^2(X)$ is the
resolution of the  orbifold  ${\rm Sym}^2(X)$, which has the
diagonal $X$ as fix point set. The resolution replaces each point
in the fixed point set by $\mathbb{P}^2$. Simple surgery and  the
smooth fibration structure of  ${\cal C}^{(2)}$ gives hence
$e({\cal C}^{(2)})=3 (e({\rm Sym}^2(X)+(3-1)e(X))=58500,$ which by
(\ref{BPSspecial}) yields  $n^{4}_5=58500+(2\cdot 6-4)(-1200)+35
\cdot 10=49250$.

The approach becomes more difficult with the number of free points
$\delta$ and at $\delta=4$ it is currently not know how to treat
the singularities of the Hilbert scheme.

On the other hand smooth curves at the `edge' of the  Castelnuovo
bound are of no principal problem. E.g., using adjunction for a
smooth complete intersection of degree $(d_1,\ldots,d_r$) in a
(weighted) projective space $\mathbb{WCP}^n(w_1,\ldots,w_{n+1})$
in Appendix A, we calculate $\chi=(2-2g)$ and see that the degree
10 genus 16 curve is the complete intersection  $(1,2,5)$. The
moduli space is calculated by counting the independent
deformations of that complete intersection. The degree five
constraint lies on the quintic, the linear constraint has five
parameters. The identification by the $\mathbb{C}^*$ action of the
ambient $\mathbb{P}^4$ shows that these parameters lie in a
$\mathbb{P}^4$. This constraint allows to eliminate one variable
from the generic quadratic constraint  which has hence $10$
parameters and a $\mathbb{P}^9$ as moduli space. So we check in
Tab. 2 the entry $n^{16}_{10}=(-1)^{13} 5\cdot 10=-50$.

Let us discuss the upper bound on the genus at which we can
completely completely fix the $F_g$ given simply the bound
(\ref{boundarycountquintic}). We claim that this bound is  $g\le
51$. At degree $20$ there is a smooth complete intersection curve
$(1,4,5)$ of that genus. We first check that this is the curve of
maximal genus in degree $20$. The Castelnuovos bound for curves in
$\mathbb{P}^4$ shows that they have smaller genus~\cite{Harris}.
We further see from the discussion in~\cite{Harris} that for
curves in $\mathbb{P}^3$ not on quadric and a cubic, which would
have the wrong degree, the Castelnuovos's bound is saturated for
the complete intersection  $(1,4,5)$. For $g=51$
(\ref{boundarycountquintic}) indicates that the gap, constant map
contribution  and regularity at the orbifold fixes $131$ of the
$151$ unknown coefficients in  (\ref{10-14-2.16}). The vanishing
of $n^{51}_{d}=0$, $1\le d\le 19$ and the value of
$n^{51}_{20}=(-1)^{ 4+ 34}
\chi(\mathbb{P}^4)\chi(\mathbb{P}^{34}) =165$ for the Euler number
of the moduli space of the smooth curve give us the rest of the
data.

\parbox{14cm} 
{

   \begin{center}
   \mbox{
             \epsfig{file=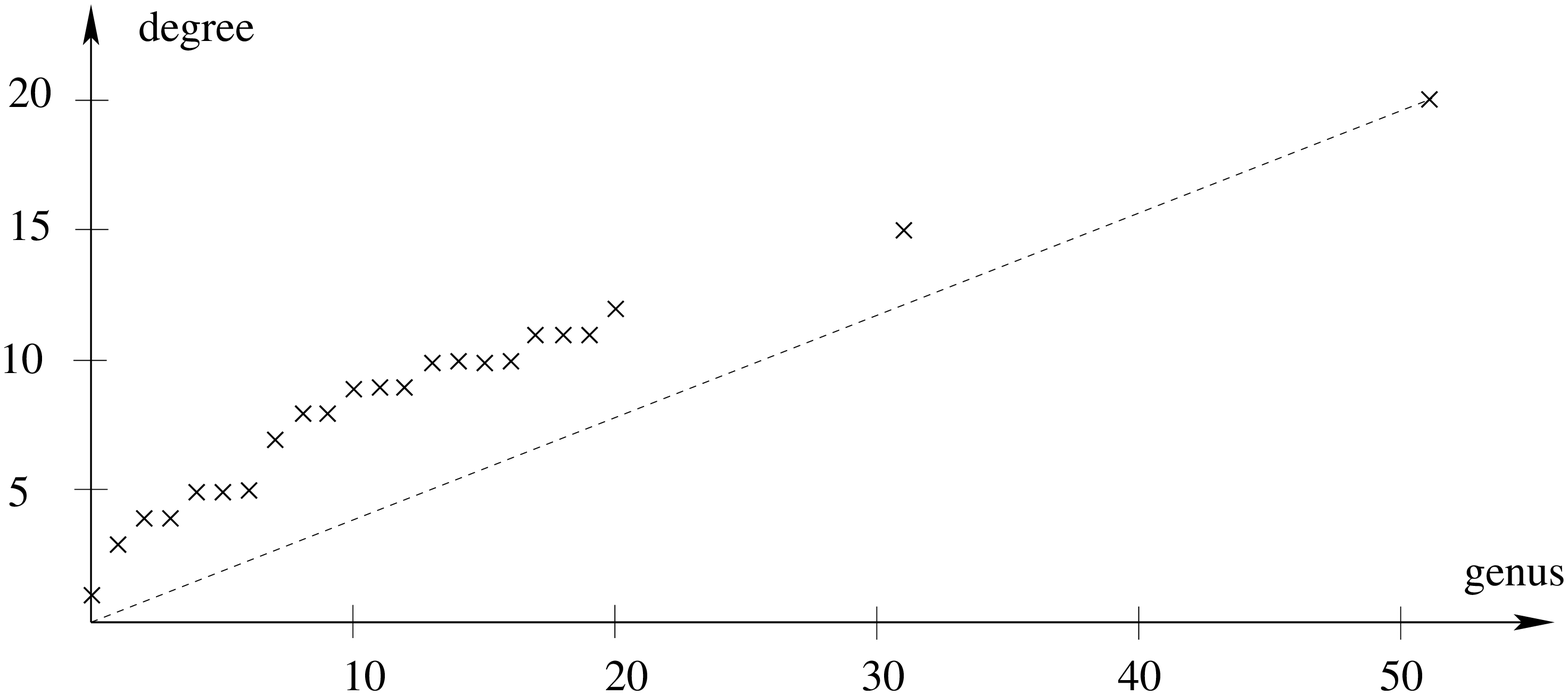,width=14cm}
}
   \end{center}
}
\vskip -2mm\noindent {\em Castelnuovo's bound for higher genus
curves on the quintic. The dashed line correspond roughly (up to taking the floor) 
to the number of coefficients in $f_g$ (\ref{10-14-2.16}) which are not fixed by
constant map contribution, conifold and orbifold boundary
conditions.}
\vskip 2mm

It is of course no problem to calculate  form the B-model the
higher genus amplitudes to arbitrary degree. For completeness we
report the first nontrivial numbers for $g=18-20$ in Tab. 3.

\begin{table}
\begin{centering}
\begin{tabular}{|r|rr|}
\hline
d &g=18 &g=19\\
\hline
\vdots& \vdots   &   \vdots \\
11& 0 &  0  \\
12&-3937166500  & -13403500  \\
13& 285683687197594125   &   -2578098061480250            \\
14&  -95076957496873268057250  &    2730012506820193210000            \\
15&  6438165666769014564325336250  &            -342304337102629200272769700   \\
16&  15209062594213864261318125134875  &    15209062594213864261318125134875           \\
\hline
\end{tabular}

\begin{tabular}{|r|r|}
\hline
d &g=20 \\
\hline
\vdots& \vdots \\
11& 0 \\
12& 0 \\
13& 10690009494250 \\
14& -59205862559233156250 \\
15& 15368486208424999875838025 \\
16& -1036824730393980503709247290500 \\
\hline
\end{tabular}
\caption{Some higher degree genus 18-20 BPS numbers for the quintic. Note
that we can calculate all Donaldson-Thomas invariants for $d=1,\ldots,12$
exactly.}
\end{centering}
\end{table}

\subsection{D-brane states on hypersurfaces in weighted projective space}

Similarly, for the sextic in $\mathbb{P}^4(1^4,2)$, the degree
$(1,2,6)$ complete intersection curve has genus $g=10$ and degree
$d={\prod_i d_i\over \prod_k w_k}=6$ in the weighted projective
space . Its moduli space is $\mathbb{P}^3$ for the degree one
constraint, i.e., we can eliminate $x_4$ form the quadric and the 7
coefficeints of the monomials $x_1^2, x_1x_2, x_1x_3, x_2^2, x_2
x_3 ,x_3^2,x_5$ form a $\mathbb{P}^6$. This yields
$n^{10}_6=(-1)^9 4 \cdot 7=-28$.

There are further checks for the octic in $\mathbb{P}^4(1^4,4)$ BPS
invariants. The complete intersection $(1^2,8)$
has total degree $2$ and genus $g=3$. The two linear constraints
describe a $\mathbb{P}^1$ in $\mathbb{P}^3$ i.e. an
$\mathbb{G}(1,3)$ with Euler number $6$ and dimension $4$, which
yields $n^3_2=6$. Similarly we have a $g=7$ complete intersection
$(1,2,8)$ of degree $4$, whose moduli space is  $\mathbb{P}^3$
times $\mathbb{P}^5$ hence $n^7_4=24$.

For the degree 10 hypersurface in $\mathbb{P}^4(1^3,2,5)$ we check the BPS invariants:
{}From the degree $(1,1,10)$ hypersuface of degree $1$ complete intersection with $g=2$. The moduli
space of the linear constraints are just the one of  point in
$\mathbb{P}^2$, i.e.$\mathbb{P}^2$, hence $n^2_1=3$. The degree
$(1,2,10)$ complete intersection with total degree $2$ and genus
$4$ has the moduli space of the linear constraint, which is
$\mathbb{P}^2$ and of the quadratic constraint is $\mathbb{P}^3$
(from the coefficients of the monomials $x_1^2,x_1 x_2,
x^2_2,x_4$), yielding $n_2^4=- 12$. Finally  the $(1,3,10)$
complete intersection with genus $7$ and degree $3$, has a moduli
space $\mathbb{P}^2$ times $\mathbb{P}^5$ (from the coefficients
of the monomials $x_1^3,x_1^2 x_2, x_1 x^2_2,x_1 x_4,x_2^3 ,x_2
x_4$) and $n_3^7=- 18$.

These checks in different geometrical situations
establish quite impressively the universality
of the gap structure at the conifold expansion.

\subsection{D-branes on complete intersections}
Here we summarize our results on one modulus complete
intersections in (weighted) projective space. More complete
results are available in \cite{webpage}. Again we can check many
BPS invariants associated to the smooth curves.

Let us check e.g.in table 7 the $n_9^{10}=15$. According
to (\ref{chi}) we see that at degree $9$ there is a smooth genus
$10$ curve, given by a complete intersection of multi degree $(1^2,3^2)$ in
$\mathbb{P}^5$. Their moduli space is the Grassmannian
$\mathbb{G}(3,5)$, which has Euler number $15$ and dimension $8$,
hence  $n_9^{10}=15$. In a very  similar way it can be seen that
the $n_8^9$ comes form a complete intersection curve of degree
 $(4,2,1^2)$ with the same moduli
space, so $n_8^9=15$ in table 8. Grassmannians related to complete
intersection are also identified with the moduli spaces of the
following smooth curves: The total degree six curve of genus seven
in table 11 is a CI of multi degree $(1^2,3,4)$. Its moduli space
is a $\mathbb{G}(2,4)$ explaining $n_6^7=10$. The degree four
curve of genus five in table 12 is of multi degree $(1^2,4^2)$ and
has moduli space $\mathbb{G}(1,3)$ yielding $n_4^5=6$.  The degree
four curve of genus five in table 13 is of multi degree
$(1^2,2,6)$ and has moduli space $\mathbb{G}(2,4)$ yielding
$n_4^5=10$.  The degree two curve of genus three in table 14 is of
multi degree $(1^2,4,6)$ and has moduli space
$\mathbb{G}(0,2)=\mathbb{P}^2$ yielding $n_2^3=3$. The moduli
space of the degree $4$ genus seven curve $(1,2,4,6)$ is an
$\mathbb{P}^2$ times the moduli space $\mathbb{P}^4$ of quadrics
in $\mathbb{WCP}^3(1^2,2^2)$, so that $n_4^7=(-1)^6 3\cdot 5=15$.
For the degree $(6,6)$ complete intersection in
$\mathbb{WCP}^3(1^2,2^2,3^2)$, see table 15, we have a degree one
genus two intersection $(1^2,6^2)$, whose moduli space is a point
hence $n_1^2=1$, a degree two genus four intersection $(1,2,6^2)$,
whose moduli space is $\mathbb{P}^1$ times the moduli space
$\mathbb{P}^2$ of quadrics in $\mathbb{WCP}^2(1,2^2)$ hence
$n_2^4=-6$, a degree three  genus seven intersection $(1,3,6^2)$,
whose moduli space is $\mathbb{P}^1$ times the moduli space
$\mathbb{P}^4$ of cubics in $\mathbb{WCP}^4(1,2^2,3^2)$ hence
$n_3^7=-10$ and finally  a degree four genus eleven intersection
$(1,4,6^2)$, whose moduli space is  $\mathbb{P}^1$ times the
moduli space $\mathbb{P}^5$ of quadrics in
$\mathbb{WCP}^4(1,2^2,3^2)$ hence $n_4^{11}=12$.

There are many further checks that are somewhat harder to perform.
E.g. we notice that there is a genus one degree three curve in the
$(3,2,2)$ CI in $\mathbb{P}^6$, which comes from a complete
intersection $(1^4,3)$. Now the moduli space of this complete
intersection in  $\mathbb{P}^6$ is $\mathbb{G}(2,6)$. However not
all $\mathbb{P}^2$ parametrized by $\mathbb{G}(2,6)$, which
contain the cubic, are actually in the two quadrics of the
$(3,2,2)$ CICY. We can restrict to those $\mathbb{P}^2$, which
fulfill these constraints, by considering the simultaneous zeros
of sections of two rank six bundles of quadratic forms on the
moving $\mathbb{P}^2$. These are a number of points, which is
calculated by the integral of the product of the Chern classes of
these rank 6 bundles over  $\mathbb{G}(2,6)$. Indeed we obtain,
for example with ``Schubert'' \cite{Schubert}
\begin{equation}
n_3^1=(-1)^0 \int_{\mathbb{G}(2,6)} c^2_6({\rm Sym}(2,Q))=64,
\end{equation}
which confirms the corresponding entry in table 9.

\section{Conclusions}

In this paper we solve the topological string B-model on compact
Calabi-Yau $M$ using the modularity of the $F_g$,
the wave function transformation property of $Z$ and the boundary
information imposed by effective action considerations. The method
pushes the calculation to unprecedented high genus amplitudes.
E.g. for the quintic the boundary condition count
(\ref{boundarycountquintic}) together with the simplest vanishing
arguments at large volume fixes the amplitudes up to genus $g=51$.
Beyond that the  prime mathematical problem to overcome in this
region of the moduli space is understand the degeneration of more
then four points in the relative Hilbert scheme of the universal
curves in a threefold\footnote{As a motivation and check for the
task to develop the  theory of Hilbert schemes for 3folds we
calcultated the invariants explicitly to high genus. For the
quintic to genus 20 and for all other the results up genus 12 are
available at \cite{webpage}.}. Similar problems have been
encountered in~\cite{Gaiotto:2006wm}, where it was suggested to
fix a very similar ambiguity to a anholomorphic ${\rm
SL}(2,\mathbb{Z})$-modular elliptic index of a $D4-D2-D0$ brane
system~\cite{Gaiotto:2006wm,deBoer:2006vg,DennefMoore}. There one  uses
${\rm SL}(2,\mathbb{Z})$ invariance of the index and a dual dilute gas
approximation in $AdS_3\times S^2\times M$ to fix the coeffcients 
of the ring of modular forms. The construction of the moduli-space of the  
$D4-D2-D0$ brane system uses rational GW invariants and implies non-trivial 
relations among them~\cite{Gaiotto:2006wm}. Such considerations 
could in principal provide further boundary conditions 
at large radius.

Our sharpest tool is the global control of $Z$ over ${\cal
M}(M)$ and we  expect that by a closer analysis of the RR-spectrum
at the orbifold of compact Calabi-Yau, we will be able to recover
at least the $\lceil \frac{2}{d}(g-1)\rceil$ conditions that one
loses relative to the local cases~\cite{Huang2} and solve the
model completely. We obtained not only the Gromov-Witten, the
Donaldson-Thomas and Gopakumar-Vafa invariants at infinity, but
also the local expansion at the conifold, the Gepner point and
other more exotic singularities with one or more massless states.
The leading singular terms in the effective action reflect the
massless states. The branch locus of the 13 parameter 
models has an intriguing variety of such light spectra and 
we can learn from the effective action about the singularity 
and vice versa. Stability properties of theses states have 
been analysed in App. \ref{appendixD-12-06}.

Most importantly our exact expansions do contain further
detailed information of the towers massive RR-states at these
points. We described them in natural local variables. The 
information from different genera should be of great value for the
study of stable even D-brane bound states on compact Calabi-Yau as
it is the content of the supersymmetric index of~\cite{GVII},
which is protected under deformations of the complex structure.
Non-compact Calabi-Yau such as the resolution of
$\mathbb{C}^n/G$, with $G\in {\rm SL}(n,\mathbb{C})$ have
no complex moduli. The issue does not arise and the situation is
better understood, see e.g. \cite{Bridgeland,Aspinwall} for reviews.

One can also use the  explicit expansions  to study  the
integrable theories that have been associated to the local
expansion of the topological string on Calabi-Yau manifold, such
as the $c=1$ string at the conifold or the quiver gauge theories
at the orbifold, matrix models e.g. at the ADE singularities and new 
ones for more exotic singularities such as the  
branch points of the complete intersections Calabi-Yau 
manifolds that we discussed here.

The ability to obtain the imprint of the BPS spectrum on the
effective action everywhere on the moduli space is of
phenomenological interest as flux compactifications drive the
theory to attractor points inside the moduli space.

Our expression are governed by the representation of the modular
group of the Calabi-Yau on almost holomorphic forms, which we
explicitly constructed from the periods, without having much of an
independent theory about them. The simpler case of the torus
suggest that such forms and their extensions should play a role
in the study of of virtually any physical amplitude --- open or closed--- 
in compactifications on the Calabi-Yau space, even as conjectured 
in the hypermultiplet sector~\cite{Rocek:2005ij}.

One may finally wonder whether the topological string B-model is
an integrable theory that is genuinely associated to this new and
barely explored class of modular forms on Calabi-Yau spaces moduli
spaces, whereas most known integrable models are associated to
abelian varieties. As it was noted
in~\cite{Verlinde:2004ck,Aganagic:2006wq, Gunaydin:2006bz}~in the
complex moduli space extended by the dilaton, called extended
phase space, one has one has rigid special K\"ahler geometry and
many aspects of the sympletic transformations and its metaplectic
realization are easier understood in the extended phase space.
There are two maps $\Phi^{(i)}: {\cal M}\rightarrow T^{(k)}_{IJ}$,
$I,J=1,\ldots,\frac{h^3}{2}$ from the complex moduli space to
tensors in the extended phase space on which
$\left(\begin{array}{cc} A&B\\ C&D \end{array}\right)\in {\rm Sp}
(h^3,\mathbb{Z})$ acts projectively like $T^{(k)}\mapsto (A
T^{(k)}+B)(C T^{k}+D)^{-1}$. For the holomorphic object
$\tau_{IJ}=\partial_I \partial_J F^{(0)}=:T^{(1)}_{IJ}(t)$, which is 
mostly discussed in this
context of the metaplectic transfomations~\cite{Verlinde:2004ck,
Aganagic:2006wq,Gunaydin:2006bz}, ${\rm Im}(\tau)$ is indefinite,
while for the non-holomorphic object ${\cal N}_{IJ}={\bar
\tau}_{IJ}+ 2 i \frac{{\rm Im}\tau_{IK} X^K {\rm Im}\tau_{IL}
X^L}{X^L{\rm Im}\tau_{KL} X^L} =:T^{(2)}_{IJ}(t,\bar t)$ comes
from the kinetic term in the 10d action whose reduction involves
the Hodge-star on $M$. It's imaginary  part ${\rm Im}({\cal N})>0$
is the kinetic term for the vector multiplets and is  hence
positive definite. In other words ${\rm Im}\Phi^{(2)}$ defines a
map to the Siegel upper space. $\Phi^{(2)}$ should relate Siegel
modular forms for admittedly very exotic subgroups~\cite{cyy} of
${\rm Sp}(4,\mathbb{Z})$ to Calabi-Yau amplitudes. Such Siegel
modular forms for abelian varieties  are also associated to $N=2$
Seiberg-Witten (gauge) theories, while the modular forms on
Calabi-Yau studied in this paper underline $N=2$ exact terms in
$N=2$ supergravity. The map $\Phi^{(2)}$ could be a manifestation
of a gravity-gauge theory correspondence for 4d theories with
$N=2$ supersymmetry.

It is no principal problem to generalize this to multi-moduli Calabi-Yau as
long as the Picard-Fuchs equations are known. These have different, more
general singularities with interesting local effective actions. In
K3 fibrations which have at least two moduli, the modular properties are much
better understood and in fact the ambiguity in the fiber is
complete fixed heterotic string calculations. Moreover these cases
have $N=2$ field theory limits, which contain further information,
which might be sufficient to solve these models \cite{GKMW}.


\vspace{0.2in} {\leftline {\bf Acknowledgments:}}

We thank M.~Aganagic, V.~Bouchard, T.~Grimm, S.~Katz, M.~Kontsevich, M.~Marino, C.~Vafa,
S.~T.~Yau and D.~Zagier for discussions. Sheldon Katz helped us with
the verifications of the BPS numbers and Cumrun Vafa with remarks
on the draft. Don Zagier's comments on~\cite{Huang} triggered many ideas here.
We thank the MSRI in Berkeley and AK thanks in particular the
Simons Professorship Program. MH/AK thank also the Simons Workshops
in Mathematics and Physics 05/06 for its hospitality.

\appendix

\section{Appendices}
\subsection{Classical intersection calculations using the adjunction
formula}
\label{intersection}

The adjunction formula\footnote{See \cite{fulton} for a
pedagogical account of these matters.} for the total Chern class
of a  for dimension $m=n-r$ smooth complete intersections  $M$ of
multi degree  $d_1,\ldots,d_r$ in a weighted projective space
$\mathbb{WCP}^n(w_1,\ldots,w_{n+1})$ is
\begin{equation}
c(T_M)=\sum_i c_i(T_M)=
{c(T_{\mathbb{WCP}})\over c({\cal N})}= {\prod_{i=1}^{n+1} (1+w_i K)\over \prod_{k=1}^r( 1+ d_k K)}=\sum_i c_i K^i\ ,
\label{adjunction}
\end{equation}
where   $c(T_{\mathbb{WCP}})=\sum_i c_i(T_{\mathbb{WCP}})=\prod_{i=1}^{n+1} (1+w_i K)$
is the total Chern class  of the weighted projective space, $K$ is its K\"ahler class and
$c({\cal N})=\prod_{k=1}^r( 1+ d_k K)$ is  the  total Chern class of the normal bundle.

Integration of a top form $\omega=x J^m$ with $J=K|_M$ over $M$ is obtained by integration
along the normal direction as
\begin{equation}
\int_{M} \omega=\int_{\mathbb{WCP}} \omega \wedge c_r({\cal N}) = {x\over \prod_{k=1}^{n+1} w_k} \prod_{k=1}^r d_k \ .
\label{integrationalg}
\end{equation}
Here we used the normalization  $\int_{\mathbb{WCP}} K^n= {1\over \prod_{k=1}^{n+1} w_k}$.
This  yields the first line below:
\begin{eqnarray}
\kappa&=&\int_{M} J^m={\prod_{k=1}^r d_k\over \prod_{i=1}^{r+1} w_i} \label{kappa}\\
\chi  &=& \int_M c_3(T_M)={c_3\over \prod_{k=1}^{n+1} w_k} \prod_{k=1}^r d_k \label{chi}\\
  c  &=& {1\over 24} \int_M c_2\wedge J={1\over 24} {c_2\over \prod_{k=1}^{n+1} w_k} \prod_{k=1}^r d_k \label{cconst}\\
  a  &=& {1\over 2}  \int_M i_* c_1(D)\wedge J= {1 \over 2}
\int_{\mathbb{WCP}} {c(T_M)\over (1+J)}\wedge J^{r+1}={{\rm coeff} \left({c(T_M)\over (1+J)},J^{m-1}\right)
\over 2\cdot \prod_{k=1}^{n+1} w_k }\label{aconst}
\end{eqnarray}
Combining (\ref{adjunction},\ref{integrationalg}) one gets the line 2
and 3. The leading $t$ terms in $F_0$ can be obtained by
calculating $Z(M)$  using (\ref{dbranecharge}), while the last
line follows from the calculation of $Z(D)$ assuming that the
$D_4$-brane is supported on $D$ the restriction of the hyperplane
class\footnote{$a$ is physically less relevant, as it does not
affect the effective action. Its value $a={11\over 2}$ obtained
for the quintic from  (\ref{aconst}) checks
with~\cite{Candelas:1990rm}} of ${\mathbb{WCP}}$ to $M$ and the Gysins
formula for smooth embeddings \cite{fulton}.

\subsection{Tables of Gopakumar-Vafa invariants}

We list the tables of BPS invariants for all the Calabi-Yau models
computed in this paper.


\begin{table}
\begin{centering}
\begin{tabular}{|r|rrrrr|}
\hline
g &d=1 &d=2 &d=3 &d=4 &d=5  \\
\hline
\, 0& \, 7884&6028452& \, 11900417220& \, 34600752005688& \,  24595034333130080\\
1& 0   &   7884&   145114704&  1773044322885&  17144900584158168\\
2& 0   &     0&        17496&    10801446444&    571861298748384\\
3& 0   &     0&          576&      -14966100&      1985113680408\\
4& 0   &     0&            6&       -47304  &       -21559102992\\
5& 0   &     0&            0&              0&           22232340\\
6& 0   &     0&            0&              0&              63072\\
7& 0   &     0&            0&              0&                  0\\
\hline
\end{tabular}
\vskip  5 pt
\begin{tabular}{|r|rrr|}
\hline
g &d=6 & d=7 &d=8 \\
\hline
0& \!\!\! 513797193321737210316&  \!\!\!\!\!\!\!\!\!\!\!2326721904320912944749252&
\!\!\!\!\!\!\!\!\!\!\!11284058913384803271372834984\\
1& \!\!\! 147664736456952923604& \!\!\!\!\!\!\!\!\!\!\!1197243574587406496495592&
\!\!\!\!\!\!\!\!\!\!\!9381487423491392389034886369\\
2&  \!\!\! 13753100019804005556&  \!\!\!\!\!\!\!\!\!\!\! 233127389355701229349884&
\!\!\!\!\!\!\!\!\!\!3246006977306701566424657380\\
3&    411536108778637626&   19655355035815833642912& 561664896373523768591774196\\
4&      1094535956564124&     628760082868148062854& 48641994707889298118864544\\
5&       -18316495265688&       3229011644338336680& 1863506489926528403683191\\
6&          207237771936&        -18261998133124302& 20758968356323626025164\\
7&            -583398600&           513634614205788& 10040615124628834206\\
8&               -146988&            -8041642037676& 1129699628821681740\\
9&                 -3168&               54521267292& -38940584273866593\\
10&                  -28&                 -43329384&  904511824896888\\
11&                    0&                  -110376& -12434437188576\\
12&                    0&                     0 &     76595605884 \\
\hline
\end{tabular}
\caption{BPS invariants $n^g_d$ on the Sextic hypersurface in $\mathbb{P}^4(1^4,2)$.}
\end{centering}
\end{table}

\begin{table}
\begin{centering}
\begin{tabular}{|r|rrrrr|}
\hline
g &d=1 &d=2 &d=3 &d=4 &d=5  \\
\hline
0& 29504& 128834912&   1423720546880& 23193056024793312&467876474625249316800 \\
1& 0   &      41312&     21464350592&  1805292092705856&101424054914016355712 \\
2& 0   &        864&       -16551744&    12499667277744&  5401493537244872896 \\
3& 0   &          6&         -177024&     -174859503824&    20584473699930496 \\
4& 0   &          0&               0&         396215800&     -674562224718848 \\
5& 0   &          0&               0&            301450&       12063928269056 \\
6& 0   &          0&               0&              4152&         -86307810432 \\
7& 0   &          0&               0&                24&             37529088 \\
8& 0   &          0&               0&                 0&               354048 \\
$\vdots$&$\vdots$&$\vdots$&$\vdots$& $\vdots$& \\
\hline
\end{tabular}
\caption{BPS invariants $n^g_d$ on the Octic hypersurface in $\mathbb{P}^4(1^4,4)$.}
\end{centering}
\end{table}

\begin{table}
\begin{centering}
\begin{tabular}{|r|rrrr|}
\hline
g &d=1 &d=2 &d=3 &d=4  \\
\hline
0& 231200& 12215785600& 1700894366474400& 350154658851324656000
\\
1&    280&   207680960&  161279120326840& 103038403740897786400
\\
2&      3&    -537976 &    1264588024791&   8495973047204168640
\\
3&      0&       -1656&     -46669244594&     61218893443516800
\\
4& 0     &         -12&        630052679&     -2460869494476896
\\
5& 0     &           0&         -1057570&       145198012290472
\\
6& 0     &           0&            -2646&        -5611087226688
\\
7& 0     &           0&              -18&         125509540304
\\
8& 0     &           0&                0&         -1268283512
\\
$\vdots$&$\vdots$&$\vdots$&$\vdots$&$\vdots$ \\
\hline
\end{tabular}
\caption{BPS invariants $n^g_d$ on the degree 10  hypersurface in $\mathbb{P}^4(1^3,2,5)$.}
\end{centering}
\end{table}

\begin{table}
\begin{centering}

\begin{tabular}{|r|rrrrrr|}
\hline
g &d=1 &d=2 &d=3 &d=4 &d=5 &d=6 \\
\hline
\, 0& 1053 & 52812 & 6424326  & 1139448384 & 249787892583& 62660964509532\\
1& 0   &  0 & 3402 &5520393  & 4820744484&3163476682080\\
2& 0   &  0 & 0 & 0 &  5520393& 23395810338\\
3& 0   &  0 &  0&  0& 0& 6852978 \\
4& 0   &  0 &  0&  0& 0&  10206\\
5& 0   &  0 &  0&  0& 0&  0 \\
\hline
\end{tabular}
\begin{tabular}{|r|rrr|}
\hline
g &d=7 &d=8 &d=9  \\
\hline
0& 17256453900822009 & 5088842568426162960  & 1581250717976557887945 \\
1&  1798399482469092 &  944929890853230501  &  473725044069553679454 \\
2&   42200615912499  &   50349477671013600  &   47431893998882182563 \\
3&      174007524240 &     785786604262830  &    1789615720312984368 \\
4&          -484542  &       2028116431098  &      21692992151427138\\
5&           158436 &           -784819773  &         36760497856020\\
6&                0 &               372762  &           -61753761036\\
7&                0 &               6318    &             -5412348\\
8&                0 &                0      &                39033\\
9&                0 &                0      &                1170\\
10&               0 &                0      &                 15\\
11      &         0 &                0      &                  0\\
\hline
\end{tabular}
\caption{$n^g_d$ for the degree  (3,3) complete intersection in $\mathbb{P}^5$.}
\end{centering}
\end{table}

\begin{table}
\begin{centering}
\begin{tabular}{|r|rrrrrr|}
\hline
g &d=1 &d=2 &d=3 &d=4 &d=5 &d=6\\
\hline
\, 0& 1280& 92288& 15655168& 3883902528& 1190923282176& 417874605342336  \\
1& 0&0& 0& -672& 16069888& 174937485184\\
2& 0& 0& 0& -8& 7680& 12679552\\
3& 0& 0& 0& 0& 0& 276864\\
4& 0& 0& 0& 0& 0& 0\\
\hline
\end{tabular}
\begin{tabular}{|r|rrr|}
\hline
g &d=7 &d=8 &d=9\\
\hline
0& \! 160964588281789696&  \!\! 66392895625625639488&  \!\! 28855060316616488359936 \\
1&  19078577926517760& 14088192680381290336& 9895851364631438617600\\
2&  494602061689344& 853657285175383648& 1137794220513866498304\\
3&  2016330670592& 14859083841009280& 49286012311292922368\\
4&  -285585152& 37334304102560& 679351051885623552\\
5& 591360& -46434384200& 1103462757073920\\
6& 7680& -8285120& -4031209095680\\
7&0& 67208& 370290688 \\
8& 0& 1520& -2270720 \\
9& 0& 15& -25600 \\
10&0& 0& 0\\
\hline
\end{tabular}

\caption{$n^g_d$ for the degree (4,2) complete intersection in $\mathbb{P}^5$.}
\end{centering}
\end{table}

\begin{table}
\begin{centering}
\begin{tabular}{|r|rrrrrr|}
\hline
g &d=1 &d=2 &d=3 &d=4 &d=5 &d=6\\
\hline
0& 720& 22428& 1611504& 168199200& 21676931712& 3195557904564  \\
1& 0& 0& 64& 265113& 198087264& 89191835056\\
2& 0& 0& 0& 0& 10080& 180870120\\
3& 0& 0& 0& 0& 0& -3696  \\
4& 0& 0& 0& 0& 0& -56 \\
5& 0& 0& 0& 0& 0& 0 \\
\hline
\end{tabular}
\begin{tabular}{|r|rrr|}
\hline
g &d=7 &d=8 &d=9\\
\hline
0& \,\, 517064870788848& \, 89580965599606752& \, 16352303769375910848  \\
1& 32343228035424&10503104916431241& 3201634967657293024 \\
2& 315217101456& 280315384261560& 178223080602086784 \\
3& 199357344& 1430336342574& 2915033921871456 \\
4& 30240& 194067288& 8888143990672 \\
5& 0& 795339& -233104896  \\
6& 0& 0& 4857552 \\
7& 0& 0& 384 \\
8& 0& 0&  0\\
\hline
\end{tabular}
\caption{$n^g_d$ for the (3,2,2) complete intersection in $\mathbb{P}^6$.}
\end{centering}
\end{table}

\begin{table}
\begin{centering}
\begin{tabular}{|r|rrrrrrr|}
\hline
g &d=1 &d=2 &d=3 &d=4 &d=5 &d=6& d=7\\
\hline
0& \,\,  512&\, 9728& \,416256& \,25703936& \,1957983744&  \, 170535923200&  \,16300354777600   \\
1& 0& 0& 0& 14752& 8782848& 2672004608& 615920502784 \\
2& 0& 0& 0& 0& 0& 1427968& 2440504320\\
3& 0&0& 0& 0& 0& 0& 86016 \\
4&0& 0& 0& 0& 0& 0& 0  \\
\hline
\end{tabular}
\begin{tabular}{|r|rrrr|}
\hline
g &d=8 &d=9 &d=10 &d=11\\
\hline
0& \!\!\!\!\!  1668063096387072& \!\!\!\!\!\!\!\!\! 179845756064329728& \!\!\!\!\!\!\!\!\!20206497983891554816& \!\!\!\!\!\!\!\!\!2347339059011866069504 \\
1& \!\!\!\!\!\!\!\!\!123699143093152& \!\!\!\!\!\!\!\!\!22984995833484288&  \!\!\!\!\!\!\!\!\! 4071465816864581632&\!\!\!\!\!\!\!\!\! 698986176207439627264\\
2&  1628589698304& 702453851520512& 236803123487243776&68301097513852719616\\
3& 2403984384& 4702943495168& 4206537025629952& 2482415474680798208\\
4& -37632& 2449622016& 16316531089408& 29624281509824512\\
5& -672& 258048& 2777384448& 73818807399424\\
6& 0& 0& 4283904& 1153891840\\
7& 0& 0& 0& 26348544\\
8& 0& 0& 0& 0\\
\hline
\end{tabular}
\caption{ $n^g_d$ for the degree (2,2,2,2) complete intersection in $\mathbb{P}^7$.}
\end{centering}
\end{table}

\begin{table}
\begin{centering}
\begin{tabular}{|r|rrrrrr|}
\hline
g &d=1 &d=2 &d=3 &d=4 &d=5 &d=6 \\
\hline
0& 1944& 223560& 64754568& 27482893704& 14431471821504& 8675274727197720\\
1&  0& 27& 161248& 381704265& 638555324400& 891094220317561\\
2& 0& 0& 0& 227448& 3896917776& 20929151321496\\
3& 0& 0& 0& 81& 155520& 75047188236\\
4& 0& 0& 0& 0& 5832& -40006768\\
5& 0& 0& 0& 0& 0& 26757\\
6& 0& 0& 0& 0& 0& 816\\
7& 0& 0& 0& 0& 0& 10\\
8&  0& 0& 0& 0& 0& 0\\
\hline
\end{tabular}
\caption{ $n^g_d$ for the degree (4,3) complete intersection in $\mathbb{WCP}^5(1^5,2)$.}
\end{centering}
\end{table}

\begin{table}
\begin{centering}
\begin{tabular}{|r|rrrrrr|}
\hline
g &d=1 &d=2 &d=3 &d=4 &d=5 &d=6 \\
\hline
0& 3712& 982464& 683478144& 699999511744& 887939257620352& 1289954523115535040\\
1& 0& 1408& 6953728& 26841854688& 88647278203648& 266969312909257728\\
2& 0& 0& 3712& 148208928& 2161190443904& 17551821510538560\\
3& 0& 0& 0& -12432& 7282971392& 362668189458048\\
4&0& 0&0& 384& -14802048& 773557598272\\
5& 0& 0& 0& 6& -22272& -7046285440\\
6& 0& 0& 0& 0& 0& 6367872\\
7& 0& 0& 0& 0& 0& 11264\\
8& 0& 0& 0& 0& 0& 0\\
\hline
\end{tabular}
\caption{$n^g_d$ for the degree (4,4) complete intersection in $\mathbb{WCP}^5(1^4,2^2)$.}
\end{centering}
\end{table}

\begin{table}
\begin{centering}
\begin{tabular}{|r|rrrrrr|}
\hline
g &d=1 &d=2 &d=3 &d=4 &d=5 &d=6 \\
\hline
0&4992&\!\!\!\!\!\! 2388768& \!\!\!\!\!\! 2732060032&\!\!\!\!\!\! 4599616564224& \!\!\!\!\!\!9579713847066240& \!\!\!\!\!\! 22839268002374163616 \\
1&0& -504& 1228032& 79275664800& 633074010435840& 3666182351842338408 \\
2& 0& -4& 14976& -13098688& 3921835430016& 128614837503143532\\
3& 0& 0& 0& 87376& -5731751168& 482407033529880\\
4&0& 0& 0& 1456& -7098624& -3978452463012 \\
5&0& 0& 0& 10& -59904& 1776341072 \\
6& 0& 0& 0& 0& 0& 18680344\\
7&0& 0& 0& 0& 0& -7176 \\
8& 0& 0& 0&  0& 0& -36\\
9&0& 0& 0& 0& 0& 0\\
\hline
\end{tabular}
\caption{ $n^g_d$ for the degree (6,2) complete intersection in $\mathbb{WCP}^5(1^5,3)$.}
\end{centering}
\end{table}

\begin{table}
\begin{centering}
\begin{tabular}{|r|rrrrr|}
\hline
g &d=1 &d=2 &d=3 &d=4 &d=5  \\
\hline
0& \ \ 15552&\ \  27904176& \ \ 133884554688&\ \  950676829466832& \ 8369111295497240640\\
1& 8& 258344& 5966034472& 126729436388624& 2512147219945401752\\
2&0& 128& 36976576& 4502079839576& 264945385369932352\\
3&0& 3& -64432& 15929894952& 9786781718701824\\
4& 0& 0& -48& -272993052& 42148996229312 \\
5& 0& 0& 0& 800065& -592538522344\\
6& 0& 0& 0& 1036& 14847229472\\
7& 0& 0& 0& 15& -148759496\\
8& 0& 0& 0& 0& 160128\\
9& 0& 0& 0& 0& 96\\
10& 0& 0& 0& 0& 0\\
\hline
\end{tabular}
\caption{$n^g_d$ for the  degree (6,4) complete intersection in
$\mathbb{WCP}^5(1^3,2^2,3)$.}
\end{centering}
\end{table}

\begin{table}
\begin{centering}
\begin{tabular}{|r|rrrr|}
\hline
g &d=1 &d=2 &d=3 &d=4 \\
\hline
0&67104& 847288224& 28583248229280& 1431885139218997920 \\
1& 360& 40692096& 4956204918600& 616199133098321280\\
2& 1& 291328& 254022248925& 102984983365762128 \\
3& 0& -928& 1253312442& 6925290146728800\\
4& 0& -6& -39992931& 104226246583368\\
5& 0& 0& 867414& -442845743788\\
6&0& 0& -1807& 53221926192\\
7& 0& 0& -10& -3192574724\\
8& 0& 0& 0& 111434794\\
9& 0& 0& 0& -1752454\\
10& 0& 0& 0& 3054\\
11&0& 0& 0& 12\\
12& 0& 0& 0& 0\\
\hline
\end{tabular}
\caption{$n^g_d$ for the degree (6,6) complete intersection in $\mathbb{WCP}^5(1^2,2^2,3^2)$.}
\end{centering}
\end{table}

\newpage
\subsection{Invariance of the generators under a change of the basis}
\label{sectioninvariance}
We have seen the topological strings can be written as polynomials
of the generators $v_1$ , $v_2$, $v_3$, and $X$. In the
holomorphic limit, these generators can be computed from the first
two solutions $\omega_0$, $\omega_1$ of the Picard-Fuchs equation.
In the followings we prove that under an arbitrary linear change
of basis of $\omega_0$ and $\omega_1$, these generators and
therefore the topological strings are actually invariant. This is
true anywhere in the moduli space. In particular, this partly
explains why the gap structure in the conifold expansion is not
affected by a change of basis of $\omega_0$ as we observed in
all cases.

Since $X=\frac{1}{1-\psi}$ is independent of the basis $\omega_0$
and $\omega_1$ , it is trivially invariant. In the holomorphic
limit, The Kahler potential and metric go like $e^{-K}\sim
\omega_0$ and $G_{\psi\bar{\psi}}\sim
\partial_{\psi}t$, where $t=\frac{\omega_1}{\omega_0}$ is the mirror
map. The generators $u$ and $v_i$ are related to $A_i$ and $B_i$,
which we recall were defined as
\begin{eqnarray} &&A_p:=\frac{(\psi\partial\psi)^p
G_{\psi\bar{\psi}}}{G_{\psi\bar{\psi}}},~~B_p:=\frac{(\psi\partial\psi)^p
e^{-K}}{e^{-K}}, ~~(p=1,2,3,\cdots)
\end{eqnarray}
So a different normalization of basis $\omega_0$ of $\omega_1$, as
well as a change of basis in $\omega_1\rightarrow \omega_1+b_1
\omega_0$ obviously do not change the generators $A_i$ and $B_i$,
and therefore the generators $u$ and $v_i$ are also invariant.

We now tackle the remaining less trivial situation, namely a
change of basis in $\omega_0$ as the following
\begin{eqnarray} \label{basis-11-04-001}
\omega_0\rightarrow \tilde{\omega}_0=\omega_0+b_1\omega_1
\end{eqnarray}
where $b_1$ is an arbitrary constant. We denote the K\"ahler
potential, metric, mirror map and various generators in the new
basis by a tilde symbol. It is straightforward to relate them to
variables in the original basis. We find the following relations
for the mirror map
\begin{eqnarray}
s=\frac{\omega_1}{\tilde{\omega}_0}=\frac{t}{1+b_1t}
\nonumber \\
\partial_{\psi}s=\frac{\partial_\psi t}{(1+b_1t)^2}
\end{eqnarray}
and the generators $A$ and $B$
\begin{eqnarray} \label{ABrelation-11-04-002}
\tilde{A}&=&\frac{\psi\partial_\psi\tilde{G}_{\psi\bar{\psi}}}{\tilde{G}_{\psi\bar{\psi}}}=A-\frac{2b_1\psi\partial_\psi t}{1+b_1\psi} \nonumber \\
\tilde{B}&=&\frac{\psi\partial_\psi\tilde{\omega}_0}{\tilde{\omega}_0}=B+\frac{b_1\psi\partial_\psi
t}{1+b_1\psi}
\end{eqnarray}
So we find the generators $A$ and $B$, as well as the generator
$u=B$ are \textit{not} invariant under the change of basis
(\ref{basis-11-04-001}). However, we recall the generator $v_1$ is
defined as
\begin{eqnarray}
v_1=1+A+2B
\end{eqnarray}
Using the equations in (\ref{ABrelation-11-04-002}) we find the
generator $v_1$ is invariant, namely $\tilde{v}_1=v_1$. To see
$v_2$ and $v_3$ are invariant, we use the derivative relations
\begin{eqnarray}
\psi\partial_\psi v_1 &=& -v_1^2-2v_2-(1+r_0)X+v_1X \label{dr-11-04-0031} \\
\psi\partial_\psi v_2 &=& -v_1v_2+v_3 \label{dr-11-04-0032}
\end{eqnarray}
where $r_0$ is a constant that appears in the relation of
generator $A_2$ to lower generators. These derivative relations
are exact and independent of the choice of the basis in asymptotic
expansion. We have show that $v_1$ and $X$ in the first equations
(\ref{dr-11-04-0031}) are invariant under a change of the basis
(\ref{basis-11-04-001}), therefore the generator $v_2$ appearing
on the right hand side must be also invariant. Applying the same
logic to the second equation (\ref{dr-11-04-0032}) we find the
generator $v_3$ are also invariant.

Our proof explains why a change of basis like
(\ref{basis-11-04-001}) does not change the gap structure around
the conifold point and seems to be related to the $SL_2$ orbit
theorem of \cite{Schmidt,CKS}. Under a change of basis, the mirror map at the
conifold point is
$\tilde{t}_D=\frac{\omega_0t_D}{\tilde{\omega}_0}$, and has the
asymptotic leading behavior $\tilde{t}_D\sim t_D\sim
\mathcal{O}(\psi)$. Recall in the holomorphic limit, the conifold
expansion is
\begin{eqnarray}
F^{(g)}_{\textrm{conifold}} =
\omega_0^{2(g-1)}(\frac{1-\psi}{\psi})^{g-1}P_g(v_1,v_2,v_3,X) ,
\end{eqnarray}
As we have shown the generators $v_i$ and therefore $P_g$ are
invariant, so in the new basis
\begin{eqnarray}
\tilde{F}^{(g)}_{\textrm{conifold}}=
(\frac{\tilde{\omega}_0}{\omega_0})^{2(g-1)}
F^{(g)}_{\textrm{conifold}}=(\frac{t_D}{\tilde{t}_D})^{2(g-1)}
F^{(g)}_{\textrm{conifold}}
\end{eqnarray}
It is clear if there is a gap structure in one basis
$F^{(g)}_{\textrm{conifold}}=\frac{(-1)^{g-1}B_{2g}}{2g(2g-2)t_D^{2g-2}}+\mathcal{O}(t_D^0)$,
the same gap structure will be also present in the other basis,
\begin{eqnarray}
\tilde{F}^{(g)}_{\textrm{conifold}}=\frac{(-1)^{g-1}B_{2g}}{2g(2g-2)\tilde{t}_D^{2g-2}}+\mathcal{O}(\tilde{t}_D^0)
\end{eqnarray}
The asymptotic expansion in sub-leading terms $\mathcal{O}(t_D^0)$
and $\mathcal{O}(\tilde{t}_D^0)$ will be different and can be
computed by the relation between $t_D$ and $\tilde{t}_D$.

Around the conifold point there is another power series solution
to the Picard-Fuchs equation that goes like $\omega_2\sim
\mathcal{O}(\psi^2)$. We also observe that the gap structure is
not affected by a change of the basis
\begin{eqnarray}
\omega_0\rightarrow \omega_0+b_2\omega_2
\end{eqnarray}
It appears to be much more difficult to prove this observation,
since now the generators $v_i$ are not invariant under this change
of basis. A proof of our observation would
depend on the specific details of the polynomial $P_g$, and
probably requires a deeper conceptual understanding of the
conifold expansion. We shall leave this for future investigation.

\subsection{Symplectic basis, vanishing cycles and massless particles}
\label{appendixD-12-06}

We can study in more details the analytic continuation of the
symplectic basis of the periods to the orbifold point $\psi=0$.
For the four hypersurface cases and two other complete
intersection models $X_{4,3}(1^5,2)$ and $X_{6,2}(1^3,2^2,3)$, the
indices $a_i$ ($i=1,2,3,4$) of the Picard-Fuchs equation are not
degenerate at the orbifold point, so there are $4$ power series
with the leading behavior of $\psi^{a_i}$, and the analytic
continuation procedure is similar to the quintic case. In our
physical explanation of the singularity structure of higher genus
topological string amplitudes in these models, we claim that for
the hypersurface cases there is no stable massless charged state around
the orbifold point, whereas for models $X_{4,3}(1^5,2)$ and
$X_{6,2}(1^3,2^2,3)$ there are nearly massless charged states of
mass $m\sim \psi^{a_2-a_1}$. Since the charge and the mass of a
D-brane wrapping cycle is determined by the mirror map parameter,
which is the ratio of two symplectic periods, it is only possible
to have massless particles if there is a rational linear
combination of the periods with the leading behavior of
$\psi^{a_k}$, $k>1$.

For a complete intersection of degree $(d_1,\cdots,d_r)$ in
weighted projective space $\mathbb{WCP}^n(w_1,\cdots,w_{n+1})$
with non-degenerate indices $a_k$, the natural basis of solutions
at the orbifold point is
\begin{eqnarray}
\omega_k^{\textrm{orb}}=\psi^{a_k}\frac{\prod_{i=1}^r\Gamma(d_ia_k)}{\prod_{i=1}^{n+1}\Gamma(w_ia_k)}
\sum_{n=0}^{\infty}(c_0\psi)^n\frac{\prod_{i=1}^{n+1}\Gamma(w_i(n+a_k))}{\prod_{i=1}^r\Gamma(d_i(n+a_k))},~~~
k=1,2,3,4
\end{eqnarray}
where
$c_0=\frac{\prod_{i=1}^rd_i^{d_i}}{\prod_{i=1}^{n+1}w_i^{w_i}}$.
We can write the sum of the series as a contour integral\footnote{Further useful
properties of the periods of the one parameter models have been established in~\cite{Lazaroiu:2000jx}. }
enclosing the positive real axis and analytically continue to the negative
real axis to relate the above basis of solutions to the known
symplectic basis at infinity $\psi=\infty$. We find the following
relation, generalizing the result for quintic case,
\begin{eqnarray} \label{D.113-12-07}
\omega_k^{\textrm{orb}}&=&(2\pi
i)^4\frac{\prod_{i=1}^r\Gamma(d_ia_k)}{\prod_{i=1}^{n+1}\Gamma(w_ia_k)}\{\frac{\alpha_k
F_0}{1-\alpha_k}-\frac{\alpha_k F_1}{(1-\alpha_k)^2}
+\frac{\alpha_k[\kappa(1+\alpha_k)-2a(1-\alpha_k)
]}{2(1-\alpha_k)^3}X_1 \nonumber \\
&& +\frac{\alpha_k[12c(1-\alpha_k)^2
+\kappa(1+4\alpha_k+\alpha_k^2)]}{6(1-\alpha_k)^4}X_0
  \}
\end{eqnarray}
here $\alpha_k=\exp(2\pi ia_k)$ and $\kappa$, $c$, $a$ are from
the classical intersection calculations in Appendix
\ref{intersection}. We find for all cases the symplectic form and
Kahler potential have the same diagonal behavior as the case of
the quintic
\begin{eqnarray}
\omega=dF_k\wedge dX_k=s_1 d\omega_1^{\textrm{orb}}\wedge
d\omega_4^{\textrm{orb}}+s_2 d\omega_2^{\textrm{orb}}\wedge
d\omega_3^{\textrm{orb}}
\end{eqnarray}
and $e^{-K}=\sum_{k=1}^4 r_k \omega_k^{\textrm{orb}}
\overline{\omega_k^{\textrm{orb}}}$, for some constants $s_1, s_2$
and $r_k$.

It is straightforward to invert the transformation
(\ref{D.113-12-07}) and study the asymptotic behavior of the
geometric symplectic basis $(F_0,F_1,X_0,X_1)$ around the orbifold
point. Generically a linear combination of the symplectic periods
$(F_0,F_1,X_0,X_1)$ is proportional to the power of $\psi$ with
the lowest index, namely, for generic coefficients $c_1, c_2, c_3,
c_4$ we have
\begin{eqnarray}
c_1X_0+c_2X_1+c_3F_1+c_4F_0\sim \omega_1^{\textrm{orb}}\sim
\psi^{a_1}
\end{eqnarray}
In order for massless particles to appear at the orbifold point,
there must be a $Sp(4,\mathbb{Z})$ transformation of the
symplectic basis such that one of periods goes to zero faster that
the generic situation, namely we should have integer coefficients $n_k\in \mathbb{Z}$
satisfying
\begin{eqnarray} \label{D.115}
n_1X_0+n_2X_1+n_3F_1+n_4F_0\sim \omega_2^{\textrm{orb}}\sim
\psi^{a_k},~~~~k>1
\end{eqnarray}

We find (\ref{D.115}) is impossible for three of the hypersurface
cases $X_5(1^5)$, $X_8(1^4,4)$ and $X_{10}(1^3,2,5)$, but possible
for the Sextic hypersurface $X_6(1^4,2)$ and the complete
intersection cases $X_{4,3}(1^5,2)$ and $X_{6,2}(1^3,2^2,3)$.
Specifically, for the Sextic hypersurface $X_6(1^4,2)$, the
condition for (\ref{D.115}) with $k=2$
\begin{eqnarray}
n_1+3n_3=0, ~~~ 3n_1+4(n_2+n_4)=0
\label{allowed1}
\end{eqnarray}
whereas for the complete intersections  $X_{4,3}(1^5,2)$ and
$X_{6,2}(1^3,2^2,3)$, the conditions for (\ref{D.115}) are
\begin{eqnarray}
n_1+3n_3=0, ~~~ n_1=4n_2+8n_4,
\label{allowed2}
\end{eqnarray}
and
\begin{eqnarray}
n_1+2n_3=0, ~~~ n_2+\frac{7}{24}n_3+n_4=0.
\label{allowed3}
\end{eqnarray}
respectively.

This fact that there are massless integer charged states possible
in the models $X_{4,3}(1^5,2)$ and  $X_{4,3}(1^5,2)$ is entirely
consistent with our physical picture, which explained the singular
behaviour of the $F^{(g)}$ form the effective action point of
view. What is very interesting is the fact that the period
degeneration at the branch point of the hypersurface models,
$X_5(1^5)$,  $X_6(1^4,2)$, $X_8(1^4,4)$ and  $X_{10}(1^3,2,5)$  
is at genus zero very similar to the cases $X_{4,3}(1^5,2)$ and  $X_{6,2}(1^3,2^2,3)$.
In particular the periods of the models have no logarithmic singularities. 
The leading behaviour of the higher genus expansion, which we obtain 
from the global properties, indicates that the BPS states,  
which are possibly massless by (\ref{allowed1},\ref{allowed2},\ref{allowed3}),   
are stable in  $X_{4,3}(1^5,2)$  and  $X_{6,2}(1^3,2^2,3)$ models, 
but  not stable in the  $X_6(1^4,2)$ model.


\end{document}